\documentclass[10pt,journal,compsoc]{IEEEtran}

\usepackage{times}
\usepackage{graphicx}
\usepackage{amsmath}
\usepackage{amssymb}
\usepackage{epsfig,subfigure,epstopdf,multirow}
\usepackage{booktabs}
\usepackage[pagebackref=true,breaklinks=true,letterpaper=true,colorlinks,bookmarks=false]{hyperref}
\usepackage{url}
\newcommand{\etal}{\textit{et al.}}

\ifCLASSOPTIONcompsoc
  \usepackage[nocompress]{cite}
\else
  \usepackage{cite}
\fi
\ifCLASSINFOpdf
\else
\fi
\begin{document}
%
\title{Learning Image-adaptive 3D Lookup Tables for High Performance Photo Enhancement in Real-time}
%
%
%
%

\author{Hui~Zeng,
        Jianrui~Cai,
        Lida~Li,
        Zisheng~Cao,
        and~Lei~Zhang,~\IEEEmembership{Fellow,~IEEE}
\IEEEcompsocitemizethanks{\IEEEcompsocthanksitem H. Zeng, J. Cai L. Li and L. Zhang are with the Department of Computing, The
Hong Kong Polytechnic University, Hong Kong. (email: \{cshzeng, csjcai, cslli, cslzhang\}@comp.polyu.edu.hk).
\IEEEcompsocthanksitem Z. Cao is with the Camera Group of DJI Innovations Co., Ltd, Shenzhen, China (e-mail: zisheng.cao@dji.com).
\IEEEcompsocthanksitem This work is supported by the Hong Kong RGC RIF grant (R5001-18).}}

%
%

\markboth{Journal of \LaTeX\ Class Files,~Vol.~14, No.~8, August~2015}%
{Shell \MakeLowercase{\textit{et al.}}: Bare Demo of IEEEtran.cls for Computer Society Journals}
%



\IEEEtitleabstractindextext{%
\begin{abstract}
Recent years have witnessed the increasing popularity of learning based methods to enhance the color and tone of photos. However, many existing photo enhancement methods either deliver unsatisfactory results or consume too much computational and memory resources, hindering their application to high-resolution images (usually with more than 12 megapixels) in practice. In this paper, we learn image-adaptive 3-dimensional lookup tables (3D LUTs) to achieve fast and robust photo enhancement. 3D LUTs are widely used for manipulating color and tone of photos, but they are usually manually tuned and fixed in camera imaging pipeline or photo editing tools. We, for the first time to our best knowledge, propose to learn 3D LUTs from annotated data using pairwise or unpaired learning. More importantly, our learned 3D LUT is image-adaptive for flexible photo enhancement. We learn multiple basis 3D LUTs and a small convolutional neural network (CNN) simultaneously in an end-to-end manner. The small CNN works on the down-sampled version of the input image to predict content-dependent weights to fuse the multiple basis 3D LUTs into an image-adaptive one, which is employed to transform the color and tone of source images efficiently. Our model contains less than 600K parameters and takes less than 2 ms to process an image of 4K resolution using one Titan RTX GPU. While being highly efficient, our model also outperforms the state-of-the-art photo enhancement methods by a large margin in terms of PSNR, SSIM and a color difference metric on two publically available benchmark datasets. Code will be released at \url{https://github.com/HuiZeng/Image-Adaptive-3DLUT}.
\end{abstract}

}

\maketitle

\IEEEdisplaynontitleabstractindextext

%
\IEEEpeerreviewmaketitle

\IEEEraisesectionheading{\section{Introduction}\label{sec:introduction}}

In the digital camera imaging process, it is an indispensable step to enhance the perceptual quality of output photos by using several cascaded modules such as exposure compensation, hue/saturation adjustment, color space conversion and manipulation, tone mapping and gamma correction \cite{karaimer2016software}. These modules are often manually tuned by experienced engineers, which is very cumbersome since the results need to be evaluated in many different scenes. The images output by digital cameras may still need post-processing/retouching to further enhance their visual quality. Unfortunately, photo retouching is also a demanding and tedious task, which requires expertise in photograph and has complicated procedures when using professional image editing tools such as PhotoShop. It is highly desirable to learn an automatic photo enhancement model, which can robustly and efficiently enhance the perceptual quality of images captured under various scenes \cite{gharbi2017deep}.


The above practical demands have motivated the research of automatic photo enhancement. Bychkovsky \etal \cite{bychkovsky2011learning} made the pioneering attempt by collecting the MIT-Adobe FiveK dataset, which contains 5,000 image pairs with human-retouched groundtruth. Given the input/output image pairs, they extracted a set of handcrafted features from them, and trained a regression model to predict the user's adjustments \cite{bychkovsky2011learning,yan2014learning}. Yan \etal \cite{yan2014learning} further learned a ranking model which divides the color enhancement process into a set of sequential operations. However, photo retouching is a complex and content-dependent task. Due to the limited feature representation capability, the models learned by these methods are hard to achieve appealing enhancement results. 

Recently, benefiting from the fast development of deep learning techniques \cite{lecun2015deep}, training deep neural networks for effective photo enhancement has becoming increasingly popular \cite{yan2016automatic,gharbi2017deep,chen2018deep,park2018distort,hu2018exposure,Wang2019CVPR,kosugi2020unpaired}. These deep enhancement methods can be further divided into two categories. The first category of methods learn a dense pixel-to-pixel mapping between the input and output image pairs \cite{chen2018deep}, or predict pixel-wise transformations to enhance the input image \cite{yan2016automatic,gharbi2017deep,Wang2019CVPR}. However, the computational and memory costs of those methods are too heavy for practical applications, especially when the resolution of input images is high. For example, for an image taken by a mobile phone equipped with the Sony IMX586 CMOS sensor with 48 megapixels, 0.54 GB memory is needed to store one single input RGB image in float32 data type. It will cost hundreds of billions of FLOPs (floating point operations) and more than 20 GB memory for these pixel-wise methods to process an image of such resolution.

\begin{figure*}[t]
\centering
\subfigure{
\begin{minipage}[b]{1.0\linewidth}
\centering
\includegraphics[width=1.0\textwidth]{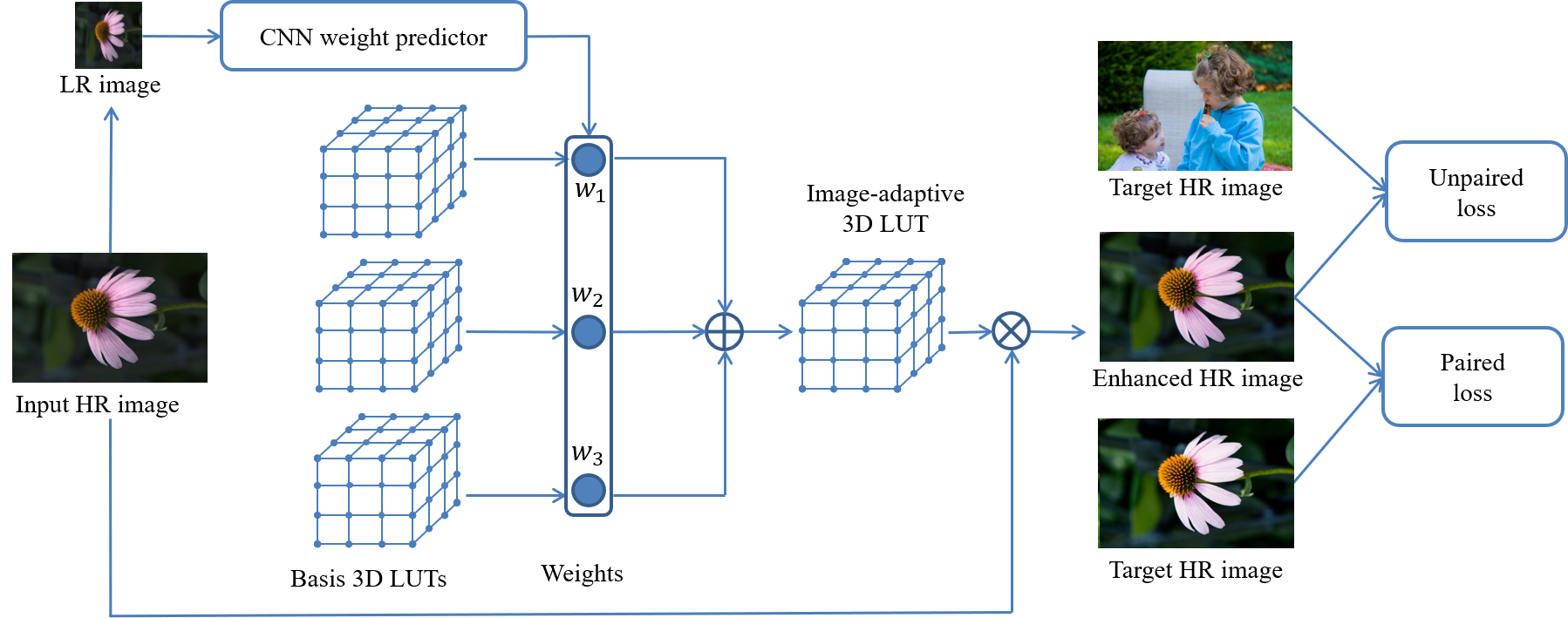}
\end{minipage}}%
\caption{Framework of the proposed image-adaptive photo enhancement method. Our method learns multiple basis 3D LUTs and a small CNN weight predictor. The CNN model works on a down-sampled version of the input image to predict content-dependent weights. The predicted weights are used to fuse the basis 3D LUTs into an image-adaptive LUT, which is then used to transform the source image. Our method can be trained using either paired or unpaired data in an end-to-end manner.}
\label{figure:framework}
\end{figure*}

Another category of deep enhancement methods train a deep neural network to predict probabilities of a set of pre-defined enhancement operators and/or parameters of a mapping curve to enhance the input image \cite{park2018distort,hu2018exposure,kosugi2020unpaired}, following the strategy of the earlier methods \cite{bychkovsky2011learning,yan2014learning}. The networks take down-sampled low-resolution (LR) images as input to predict which ones of the pre-defined operators and/or mapping curves should be used to enhance the given high-resolution (HR) source images. Compared with the first category, this category of methods have potentials to achieve higher efficiency and lower memory consumption since the network only needs to process an LR image and the pre-defined operators are usually simple. However, the employed simple operators are not able to provide adequate enhancement capability, and they are hard to be directly learned from training data. These methods \cite{park2018distort,hu2018exposure,kosugi2020unpaired} thus employ the reinforcement learning strategy to iteratively enhance the input image, which however severely sacrifices the efficiency. 

To address the limitations of existing methods on both enhancement quality and efficiency, we propose a new method which has advantages of high-efficiency, low memory-consumption and high-quality. Our method is partially motivated by the camera imaging pipeline \cite{karaimer2016software} and professional image editing tools where the 1D and 3D lookup tables (LUTs) are used to adjust the hue, saturation, exposure, color and tone of photos. As one of the most classical yet widely used photo adjustment techniques, 3D LUT can achieve stable photo enhancement performance at very high efficiency. However, those 3D LUTs are mostly manually tuned by experienced experts in a very cumbersome and costly process. Furthermore, once manually tuned, the LUTs are fixed but not adaptive to different scenes. As a result, many different modes in digital cameras and dozens of LUTs in image editing tools are preset for user's selection to fit for different scenes, which brings inconvenient user experience.

To achieve more expressive and intelligent photo enhancement capability while maintaining the advantages of 3D LUT such as high-efficiency and stability, we propose an end-to-end framework to learn image-adaptive 3D LUTs from high-quality images retouched by human experts. Fig. \ref{figure:framework} illustrates the framework of the proposed method, which learns several (e.g., 3) basis 3D LUTs and a small CNN weight predictor. The basis 3D LUTs ensure that the color transform space from input to output images can be well covered, while the CNN model predicts content-dependent weights based on the down-sampled input image. The predicted weights are used to fuse the basis 3D LUTs into an image-adaptive one, which is then used to transform the source image. This process is functionally similar to the operation of choosing different modes in digital cameras or choosing different LUTs in image editing tools but it is automatic and more flexible. The proposed adaptive LUT based photo enhancement is very fast because only the trilinear interpolation is involved and this operation can be easily parallelized using GPU \cite{LUTnvidia}. The whole model can be trained using either paired or unpaired data with different loss functions.

The highlights of this work are summarized as follows:
\begin{itemize}
\vspace{-4pt}
\item To the best of our knowledge, we are the first to learn 3D LUTs using either paired or unpaired datasets for automatic photo enhancement. More importantly, our proposed learning architecture can learn image-adaptive 3D LUTs to achieve intelligent and high performance photo enhancement, which cannot be achieved by the current 3D LUT models.
\item Our proposed model contains less than 600K parameters and takes less than 2 ms to process an image of 4K resolution using one Titan RTX GPU, which is highly efficient for deployment in practical applications.
\item Our extensive experiments on two publically available benchmark datasets validate that under either paired or unpaired settings, our model significantly outperforms state-of-the-art photo enhancement methods both quantitatively and qualitatively.
\end{itemize}

\section{Related work}

Photo enhancement is a long-standing topic with wide applications in camera imaging pipeline and photo retouching. We briefly review the related work towards these applications and the employed photo enhancement techniques as follows.

\textbf{Enhancement in camera imaging pipeline.} Camera imaging pipeline is a complex process which converts the sensor raw data to perceptually friendly color images. Under the divide and conquer principle, this process consists of a set of cascaded modules. For the enhancement part, those modules include exposure correction \cite{takahashi1998photographic,yuan2012automatic}, color constancy \cite{finlayson2004shades,gijsenij2011computational}, contrast enhancement \cite{kim1997contrast,cai2018learning},  color manipulation \cite{mukherjee2008enhancement,kim2012new} and tone mapping \cite{mantiuk2008display,lim2015systems}, each of which has been widely studied in the past. According to \cite{karaimer2016software}, most of these modules in practical systems employ a common technique, i.e., 1D or 3D LUT. These LUTs are usually manually tuned and fixed in the imaging pipeline, which are however not flexible and intelligent enough. This motivates us to learn image-adaptive 3D LUTs to improve both the expressiveness and flexibility of photo enhancement in imaging pipeline.

Considering that the use of cascaded modules in existing imaging pipeline may result in accumulated errors, several attempts have been made to learn the entire pipeline using a deep neural network \cite{nam2017modelling,schwartz2018deepisp, liang2019cameranet}. However, the robustness of such methods is hard to ensure and the computational cost is too heavy for camera devices. Different from these methods, we aim to improve the enhancement modules of the imaging pipeline, such as exposure correction, color manipulation and tone mapping, by learning image-adaptive 3D LUTs. Our method improves the flexibility and expressiveness of traditional imaging pipeline while maintaining their advantages such as stability and efficiency.

\textbf{Enhancement in photo retouching.} Photo retouching is widely used to further improve the perceptual quality of photos captured by digital cameras. Many professional software products (e.g., PhotoShop, Lightroom, 3D LUT Creator) and mobile applications (e.g., Snapseed, VSCO, Pollar) are designed for this purpose. However, using these tools to effectively render a high-quality photo requires much domain knowledge in photography and a lot of practices to get familiar with the operations. To alleviate the cost, many preset LUTs are provided in the tools or available on the internet. However, a preset LUT usually only fits for a specific type of scenes. It is still tedious and time-consuming to handle many photos with different content. Our proposed method could learn image-adaptive 3D LUTs which can automatically enhance photos based on their content.

\textbf{Example-based photo transfer/enhancement.} The \textit{example-based} methods \cite{reinhard2001color,kang2010personalization,hwang2012context,liu2014autostyle,lee2016automatic} transfer the color and style of a reference image to the target image. Traditionally, handcrafted features are designed to characterize the context and attribute of an example image, and manually designed operators are employed to perform the transformation. Recently, deep neural networks have been employed to extract more representative features and achieve more powerful and flexible (sometimes artistic) style transfer \cite{gatys2016image,luan2017deep,huang2017arbitrary}. However, it is inconvenient to find a proper reference image in practical applications. Our method differs from this type of research in that it is totally automatic after the model is trained.

\textbf{Learning-based photo enhancement.} The learning-based methods aim to learn an enhancement model from input and target image datasets. The target images are usually retouched by human experts. It is expected that the trained model can automatically enhance an input image without additional operations. The pioneering work along this line is reported in \cite{bychkovsky2011learning}, where Bychkovsky \textit{et al.} developed the MIT-Adobe FiveK dataset and learned a mapping curve to approximate photographers' adjustments. In \cite{yan2014learning}, a dataset is collected to record the manual enhancement process of each image and a ranking model is learned based on the ordered image pairs. However, both these two methods learn a shallow model with a set of handcrafted features, which have limited representation capability.

Recently, significant progress has been made by training neural networks for photo enhancement. Yan \etal \cite{yan2016automatic} fed a set of image descriptors into a multi-layer perceptron to predict pixel-wise quadratic color transforms. Gharbi \etal \cite{gharbi2017deep} proposed to learn the transformation coefficients in the bilateral space with high efficiency. Park \etal \cite{park2018distort} introduced a deep reinforcement learning strategy for step-wise image enhancement by learning to predict the discrete probabilities of a set of predefined operators. Satoshi and Toshihiko \cite{kosugi2020unpaired} employed a similar reinforcement learning strategy to learn an unpaired photo enhancement model. In addition to employing predefined operators, Hu \etal \cite{hu2018exposure} trained a neural network to fit a mapping curve, which can be viewed as a 1D LUT, to conduct photo enhancement. Chen \etal \cite{chen2018deep} employed a U-Net model to learn pixel-to-pixel mapping between input and output image pairs, while Wang \etal \cite{Wang2019CVPR} trained a model to estimate the pixel-wise illumination map of input image.

Like \cite{park2018distort,hu2018exposure,kosugi2020unpaired}, our method involves both basic operators (e.g. 3D LUTs) and CNN model for image enhancement; however, there are several significant differences. First, the operators employed in \cite{park2018distort,hu2018exposure,kosugi2020unpaired} are pre-defined and fixed, while the 3D LUTs in our method are learned from training data. Second, the pre-defined operators in previous methods, such as global exposure adjustment, contrast and saturation adjustment using one single parameter, are relatively simple and have limited expressive capability. The image-adaptive 3D LUTs learned in our method are much more expressive. Third, our method outputs the enhancement result using one single forward process, which is several orders faster than the iterative process in \cite{park2018distort,hu2018exposure,kosugi2020unpaired}.

Unsupervised learning of photo enhancement models using unpaired data has recently been studied in \cite{chen2018deep,hu2018exposure,deng2018aesthetic,kosugi2020unpaired}. Our model can also be trained on unpaired data using an adversarial loss, and it demonstrates significant advantages over previous methods in both performance and efficiency. Details can be found in the section of experiments.

\section{Methodology}

In this section, we present in detail our framework to learn image-adaptive 3D LUTs using either paired or unpaired data. 

\subsection{3D LUT and trilinear interpolation}

\begin{figure}[t]
\centering
\subfigure{
\begin{minipage}[b]{1.0\linewidth}
\centering
\includegraphics[width=1.0\textwidth]{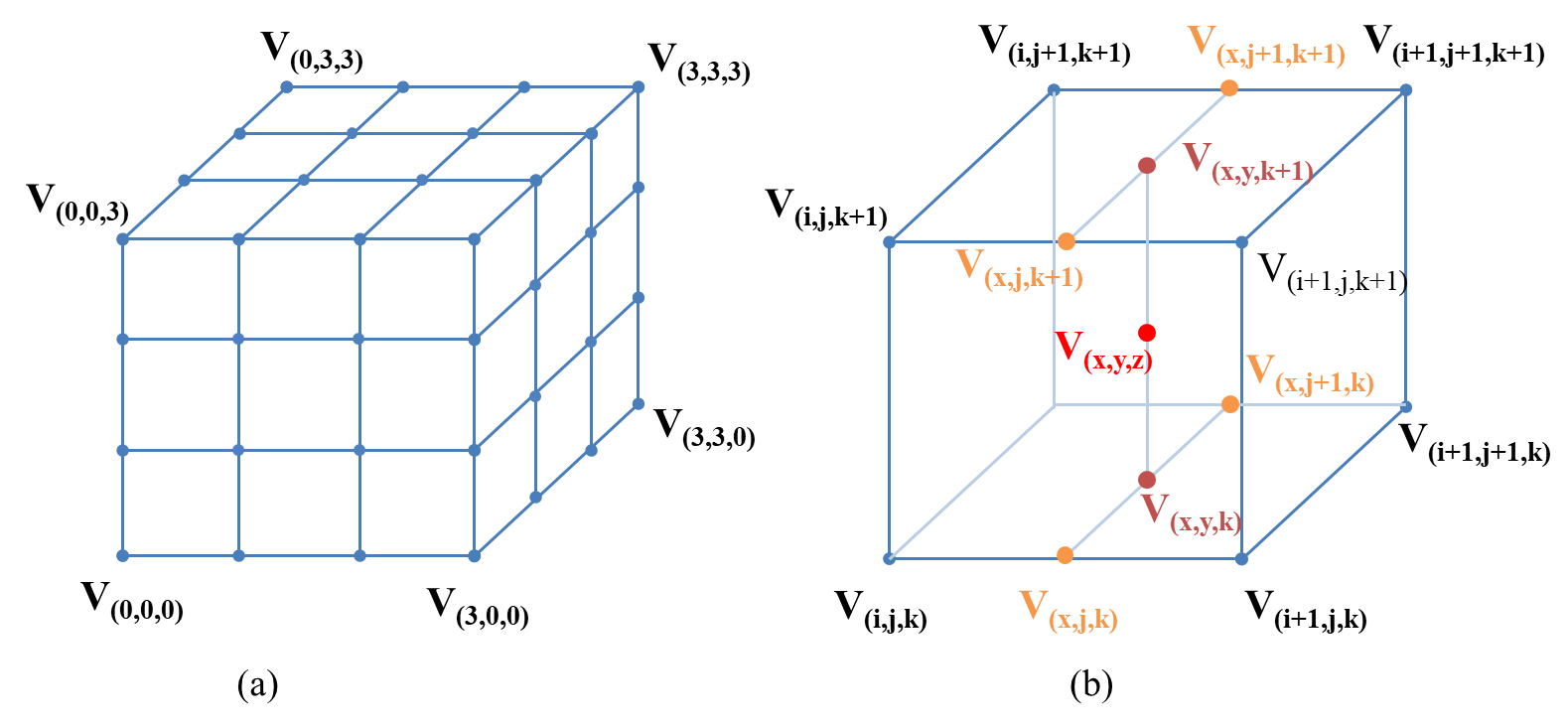}
\end{minipage}}
\caption{Illustration of (a) a 3D LUT containing $4^3$ elements and (b) the trilinear interpolation of one input.}
\label{figure:3DLUT}
\end{figure}

3D LUT is a classical yet very effective and widely used technique for photo enhancement.
As shown in Fig. \ref{figure:3DLUT}(a), a 3D LUT defines a 3D lattice which consists of $M^3$ elements $\{\mathbf{V}_{(i,j,k)}\}_{i,j,k=0,...,M-1}$, where $M$ is the number of bins in each color channel.  Each element $\mathbf{V}_{(i,j,k)}$ defines an indexing RGB color $\{r^I_{(i,j,k)},g^I_{(i,j,k)},b^I_{(i,j,k)}\}$ and the corresponding transformed output RGB color $\{r^O_{(i,j,k)},g^O_{(i,j,k)},b^O_{(i,j,k)}\}$. The precision of this transformation is determined by the setting of $M$, which is usually set to 33 in practice. Given the value of $M$, the indexing RGB colors $\{r^I_{(i,j,k)},g^I_{(i,j,k)},b^I_{(i,j,k)}\}_{i,j,k=0,...,M-1}$ are obtained by uniformly discretizing the RGB color space. Different 3D LUTs have different output RGB colors $\{r^O_{(i,j,k)},g^O_{(i,j,k)},b^O_{(i,j,k)}\}_{i,j,k=0,...,M-1}$, which are the learnable parameters in our method. When $M=33$, a 3D LUT contains 108K ($3M^3$) parameters.

The color transformation in 3D LUT is achieved by two basic operations: lookup and trilinear interpolation. Given an input RGB color $\{r^I_{(x,y,z)},g^I_{(x,y,z)},b^I_{(x,y,z)}\}$, a lookup operation is conducted to find its location $(x,y,z)$ in the 3D LUT lattice:
\begin{equation}\label{equ:lookup}
x = \frac{r^I_{(x,y,z)}}{s}, y = \frac{g^I_{(x,y,z)}}{s}, z = \frac{b^I_{(x,y,z)}}{s},
\end{equation}
where $s=\frac{C_{max}}{M}$ and $C_{max}$ is the maximum color value. After the location of the input RGB color is computed, its nearest 8 surrounding elements can be used to interpolate the output RGB via trilinear interpolation, which is illustrated in Fig. \ref{figure:3DLUT}(b). Let $i = \lfloor x\rfloor, j = \lfloor y\rfloor, k = \lfloor z\rfloor$, where $\lfloor\cdot\rfloor$ is the floor function, and let $d_x=\frac{r^I_{(x,y,z)}-r^I_{(i,j,k)}}{s}$, $d_y=\frac{g^I_{(x,y,z)}-g^I_{(i,j,k)}}{s}$, $d_z=\frac{b^I_{(x,y,z)}-b^I_{(i,j,k)}}{s}$. The transformed output RGB $\{r^O_{(x,y,z)},g^O_{(x,y,z)},b^O_{(x,y,z)}\}$ can be derived by the trilinear interpolation as follows:
\begin{align}\label{equ:trilinear}
\scriptstyle
c^O_{(x,y,z)} 
&\scriptstyle= (1-d_x)(1-d_y)(1-d_z)c^O_{(i,j,k)} + d_x(1-d_y)(1-d_z)c^O_{(i+1,j,k)} \notag\\
&\scriptstyle+ (1-d_x)d_y(1-d_z)c^O_{(i,j+1,k)} + (1-d_x)(1-d_y)d_zc^O_{(i,j,k+1)} \notag\\
&\scriptstyle+ d_xd_y(1-d_z)c^O_{(i+1,j+1,k)} + (1-d_x)d_yd_zc^O_{(i,j+1,k+1)} \notag\\
&\scriptstyle+ d_x(1-d_y)d_zc^O_{(i+1,j,k+1)} + d_xd_yd_zc^O_{(i+1,j+1,k+1)},
\end{align}
where $c \in \{r,g,b\}$. The above trilinear interpolation is sub-differentiable and it is easy to derive the gradient of $c^O_{(i,j,k)}$. Since the trilinear interpolation of each input is independent of other pixels, this transformation can be easily parallelized using GPU.

\subsection{Learning image-adaptive 3D LUTs}\label{section:image-adaptive}

The traditional 3D LUT based image enhancement methods have two major limitations. First, the 3D LUTs are mostly manually designed, which is cumbersome and costly. Second, one 3D LUT can only provide a fixed transformation, which is hard to adapt to different scenes. Though existing camera devices or image editing tools provide a set of preset LUTs which can be manually selected to achieve different enhancement effects, this solution is however not flexible and convenient enough. 


To address the first limitation, we choose to adopt the data-driven approach. For example, with a dataset which contains source and target images, we can learn one or multiple LUTs by optimizing some objective function. Once learned, those LUTs can be used to automatically enhance the given images. To address the second limitation, the model should be content-aware and adaptive to the input, and we propose to learn image-adaptive 3D LUTs to achieve this goal. An intuitive idea is to learn a classifier to perform scene classification, and then use different 3D LUTs to enhance different images. This strategy is adopted by many camera devices and image editing tools. Suppose that $N$ 3D LUTs, denoted by $\{\phi_n\}_{n=1,...,N}$, are learned. The classifier outputs $N$ probabilities $\{p_n\}_{n=1,...,N}$ for scene classification. This 3D LUT selection process can be described as follows:
\begin{equation}\label{equ:enhance}
q = \phi_i(x), \quad s.t. \quad i = \arg\max_n{p_n},
\end{equation}
where $x$ denotes an input image and $q$ denotes the output. The above hard-voting strategy has several drawbacks. First of all, it is not easy to categorize the many possible scenes into a set of well-defined classes so that each class of images share the same LUT. Second, it requires a lot of 3D LUTs to cover a variety of scenes since each 3D LUT is independently used to enhance the input image. Third, the classifier is trained independently of the 3D LUTs so that the collaboration between them is not optimal. If the scene classification is incorrect, the selected 3D LUT may lead to very bad results. 

In this work, we achieve image-adaptive 3D LUTs learning using a more effective and efficient way. We jointly learn a few basis 3D LUTs $\{\phi_n\}_{n=1,...,N}$ together with a small CNN model $f$ which predicts weights for the output of each LUT. For an input image $x$, the final enhancement output is obtained as: 
\begin{equation}\label{equ:enhance1}
q = \sum\nolimits_{n=1}^{N}w_n\phi_n(x),
\end{equation}
where $\{w_n\}_{n=1,...,N}=f(x)$ are the content-dependent weights output by the CNN model. Our design only needs several (e.g., $N=3$) basis 3D LUTs to transform the input image and employs a soft-weighting strategy to achieve image content adaptive transformation. Furthermore, considering that pixels are transformed independently using trilinear interpolation in 3D LUTs, the operations in Eq. \ref{equ:enhance1} can be simplified as:
\begin{equation}\label{equ:enhance2}
q = (\sum\nolimits_{n=1}^{N}w_n\phi_n)(x).
\end{equation}
That is, we can first fuse the multiple 3D LUTs into an image-adaptive 3D LUT, and then perform only one single trilinear interpolation. Such a design further improves the speed of our method to process high-resolution images. The objective function of our learning scheme can be written as follows:
\begin{equation}\label{equ:objective}
min_{f,\phi_n} \mathcal{L}(q,y),
\end{equation}
where $f$ and $\phi_n$ are the CNN model and basis 3D LUTs to be learned, $\mathcal{L}(q,y)$ denotes some loss functions (defined between the transformation output $q$ and target $y$) and regularization terms which will be discussed later.

\begin{table}[t]
\small
\centering
\caption{Details of the CNN weight predictor. C, D, F and $N$ represent convolutional block, dropout, fully-connected layer and number of weights, respectively.}
\label{table:classifier}
\begin{tabular}{|c|c|c|c|c|c|c|c|}
\hline
Layer  & C1 & C2 & C3 & C4 & C5  &D& F \\\hline
Input  & 256& 128& 64& 32 & 16  &8& 8\\
Kernel & 3  & 3  & 3  & 3  & 3    &-- & 8\\
Channel& 16 & 32 & 64 & 128& 128&128&$N$ \\
\hline
\end{tabular}
\end{table}

The CNN weight predictor aims to understand the global context such as brightness, color and tones of the image to output content-dependent weights. Therefore, it only needs to work on the down-sampled version of the input image to largely save computational cost. Given an input image of any resolution, we simply use bilinear interpolation to downsample it to $256\times256$ resolution for high efficiency. We have evaluated more complex downsampling strategies such as bicubic downsampling and box filter downsampling but found little difference on the final performance. The detailed architecture of the CNN model is described in Table \ref{table:classifier}. It consists of 5 convolutional blocks, one dropout layer and one fully-connected layer. Each convolutional block contains a convolutional layer, a leaky ReLU \cite{xu2015empirical} and an instance normalization \cite{ulyanov2016instance} layer. The dropout rate is set to 0.5. The fully-connected layer output $N$ weights. The whole CNN model contains only 269K parameters when $N=3$.

Although our design makes 3D LUT image-adaptive, 3D LUT itself cannot adapt to different local areas within an image, which may result in less satisfied local enhancement quality in some challenging cases, as will be discussed in Section \ref{sec:limitation}.

\subsection{Learning criteria}

According to the given training data (e.g., paired or unpaired data), we can employ different loss functions in Eq. \ref{equ:objective} to train our model. 

\textbf{Pairwise learning.} When a set of input images and their corresponding (human annotated) groundtruth images are available, one can simply use supervised learning methods to learn the photo enhancement model. Suppose that there are a number of $T$ training pairs $\{x_t, y_t\}_{t=1,2,...,T}$, where $x_t$ and $y_t$ denote a pair of input and target images. We simply employ the MSE loss to train the model:
\begin{equation}\label{equ:mse}
\mathcal{L}_{mse} = \frac{1}{T}\sum\nolimits_{t=1}^{T}\|q_t-y_t\|^2,
\end{equation}
where $q_t$ is the output of Eq. \ref{equ:enhance2} with $x_t$ as input. Note that other pairwise loss functions such as the $L_1$ loss or color difference loss can also be used. We choose MSE loss just for its simplicity. 

\textbf{Unpaired learning.} Though pairwise learning is easy to implement and tends to generate good enhancement results, the collection of pairwise data may be expensive or very difficult. Unsupervised learning of deep models using unpaired data has become increasingly popular recently, owing to the rapid development of Generative Adversarial Networks (GAN) \cite{goodfellow2014generative}. By using the GAN loss, we can also learn photo enhancement models with unpaired training data. In such case, the multiple 3D LUTs together with the CNN weight predictor can be viewed as the Generator, denoted by $G$. One more CNN is introduced as the discriminator, denoted by $D$. In our experiment, $D$ has the same architecture as the CNN weight predictor (refer to Section \ref{section:image-adaptive}) with only two minor changes: the instance normalization and dropout layers are removed and the output dimension of the FC layer is set to 1.

The generator $G$ is trained to minimize:
\begin{equation}\label{equ:generator}
\mathcal{L}_{G} = \mathop{\mathbb{E}}\limits_x[-D(G(x))]+\lambda_1 \mathop{\mathbb{E}}\limits_x[\|G(x)-x\|^2],
\end{equation}
where the term $-D(G(x))$ enforces the generator $G$ to fool the discriminator $D$, and the term $\|G(x)-x\|^2$ ensures that the output image preserves the same content as the input image. $\lambda_1$ is a constant parameter to balance the two terms. We set it as 1,000 following the setting in \cite{chen2018deep}.

The discriminator is trained to minimize:
\begin{equation}\label{equ:discriminator}
\mathcal{L}_{D} = \mathop{\mathbb{E}}\limits_x[D(G(x))]-\mathop{\mathbb{E}}\limits_y[D(y)]+\lambda_2 \mathop{\mathbb{E}}\limits_{\widehat{y}}[(\|\bigtriangledown_{\widehat{y}}D(\widehat{y})\|_2-1)^2],
\end{equation}
where the first two terms are the standard discriminator losses, the third term is the gradient penalty proposed in \cite{gulrajani2017improved} to stabilize the training process, and $\widehat{y}$ is an example linearly interpolated by a pair of points from the generator distribution and target distribution. Parameter $\lambda_2$ is fixed as 10 throughout our experiments following the setting in \cite{gulrajani2017improved}. The final GAN loss is:
\begin{equation}\label{equ:gan}
\mathcal{L}_{gan} = \mathcal{L}_{G} + \mathcal{L}_{D}.
\end{equation}
The generator and discriminator are iteratively optimized at equal pace in our experiments.

\subsection{Regularization}

Using the $\mathcal{L}_{mse}$ loss or the $\mathcal{L}_{gan}$ loss, we can train the 3D LUTs and CNN weight predictor using gradient descent algorithms such as SGD with momentum \cite{krizhevsky2012imagenet} or Adam \cite{kingma2014adam}. However, the optimized 3D LUTs may have unsmooth surfaces (please refer to Fig. \ref{figure:visualization_LUT}(a) for some visualized examples of learned 3D LUTs). The abrupt color changes in neighboring lattices of the 3D LUT may amplify the color difference after color transformation, causing some banding artifacts \cite{wiki:banding} in smooth areas of the enhanced image (please refer to Fig. \ref{figure:regularization}(a) for an example). In order to make the learned 3D LUTs more stable and robust, we introduce two regularization terms to the optimization process.


\textbf{Smooth regularization.} In order to stably convert the input RGB values into the desired color space without generating much artifacts, the output RGB values of the 3D LUTs should be locally smooth. The total variation (TV) \cite{rudin1992nonlinear} is a classical smooth regularization technique for image restoration. Inspired by TV, we propose a 3D smooth regularization term on the learning of 3D LUTs as follows:
\begin{align}\label{equ:tv1}
\mathcal{R}_{TV} &= \sum_{c \in \{r,g,b\}}\sum_{i,j,k}(\|c^O_{(i+1,j,k)}-c^O_{(i,j,k)}\|^2 + \|c^O_{(i,j+1,k)} \notag\\ &-c^O_{(i,j,k)}\|^2+\|c^O_{(i,j,k+1)}-c^O_{(i,j,k)}\|^2).
\end{align}
We choose the $L_2$ distance rather than the $L_1$ distance in the above term to achieve smoother regularization.

In addition to the 3D LUTs themselves, the content-dependent weights $w_n$ predicted by the CNN are also important to our image-adaptive LUT learning scheme. We thus introduce the $L_{2}$-norm regularization on the predicted weights to improve the smoothness of the adaptively weighted 3D LUT. The overall smooth regularization term is as follows:
\begin{align}\label{equ:tv2}
\mathcal{R}_{s} &= \mathcal{R}_{TV} + \sum_{n}\|w_{n}\|^2.
\end{align}

\textbf{Monotonicity regularization.} Apart from smoothness, monotonicity is another property 3D LUTs should possess for two reasons. First, a monotonic transformation can preserve the relative brightness and saturation of input RGB values, ensuring natural enhancement result. Second, in practice the training data may be insufficient to cover the entire color space. Ensuring monotonicity can help to update the parameters that may not be activated by input RGB values, improving the generalization capability of the learned 3D LUTs. We therefore design a monotonicity regularization on 3D LUT learning as follows:
\begin{align}\label{equ:tv3}
\mathcal{R}_{m} &= \sum_{c \in \{r,g,b\}}\sum_{i,j,k}[g(c^O_{(i,j,k)}-c^O_{(i+1,j,k)}) + g(c^O_{(i,j,k)} \notag\\
& -c^O_{(i,j+1,k)}) + g(c^O_{(i,j,k)}-c^O_{(i,j,k+1)})],
\end{align}
where $g(\cdot)$ is defined as the standard ReLU operation, i.e., $g(a)=max(0,a)$ . The monotonicity regularization ensures that the output RGB values $c^O_{(i,j,k)}$ increases with the index $i, j, k$ and larger $i, j, k$ indices corresponds to larger input RGB values in the 3D LUT lattice.

\subsection{Final training losses and implementation}
By incorporating the two regularization terms, the final loss functions used in paired learning and unpaired learning are as follows:
\begin{equation}\label{equ:paired}
\mathcal{L}_{paired} = \mathcal{L}_{mse} + \lambda_s\mathcal{R}_{s} + \lambda_m\mathcal{R}_{m},
\end{equation}
\begin{equation}\label{equ:unpaired}
\mathcal{L}_{unpaired} = \mathcal{L}_{gan} + \lambda_s\mathcal{R}_{s} + \lambda_m\mathcal{R}_{m},
\end{equation}
where the two constant parameters $\lambda_s$ and $\lambda_m$ are used to control the effects of the smooth and monotonicity regularization terms, respectively. We emperically set $\lambda_s = 0.0001$ and $\lambda_m = 10$ via an ablation study.

We implement our method based on PyTorch \cite{paszke2017automatic}. The trilinear interpolation of 3D LUT is parallelized via a CUDA parallel implementation. The standard Adam \cite{kingma2014adam} optimizer with default parameters is employed to train our model. The batch size is set to 1 and the learning rate is fixed at $1e{-4}$ and $2e{-4}$ for paired and unpaired experiments, respectively. We randomly crop image patches with scale in the range [0.6,1.0], horizontally flip input/target image pairs, slightly adjust the brightness and saturation of input images for data augmentation. Among the $N$ 3D LUTs to be learned, the first one is initialized as an identity map while the others are initialized as zero maps. The bias of the FC layer is set to 1 which makes the initially predicted weight approach to 1. Such an initialization ensures that the initial output of our model is not far away from the input and thus increases the training speed. The other parameters of the CNN weight predictor are randomly initialized as in \cite{glorot2010understanding}. 

\section{Experiments}

\subsection{Experimental setup}\label{subsection:experimental setup}

\textbf{Datasets.} We conduct experiments on two datasets: MIT-Adobe FiveK \cite{bychkovsky2011learning} and HDR+ \cite{hasinoff2016burst}. The MIT-Adobe FiveK dataset \cite{bychkovsky2011learning} is currently the largest photo enhancement dataset with human retouched groundtruth. The dataset contains 5,000 raw photos, each of which was retouched by five human experts. Following the common practice \cite{gharbi2017deep, chen2018deep, park2018distort,Wang2019CVPR}, the images retouched by expert C are used as the groundtruth in our experiments. The HDR+ dataset \cite{hasinoff2016burst} is a burst photography dataset collected by Google camera group for research of high dynamic range (HDR) and low-light imaging on mobile cameras. It consists of a total of 3,640 scenes, each of which contains 2 to 10 raw photos. We train our model on this dataset to learn the HDR enhancement module of camera imaging pipeline. Specifically, the intermediate result of the aligned and merged frames (in DNG format) are transformed into TIF images with 16-bit dynamic range as the input, and the JPG images generated by manually fine-tuned HDR imaging pipeline are used as the groundtruth. 

We conduct experiments on two different resolutions: 480p and the image original resolution (generally 12-13 megapixels). The aspect ratios of source images are mostly 4:3 or 3:4. To avoid unnatural deformation, we resize the short side of source images to 480 pixels for experiments on the 480p resolution. 

\textbf{Application settings.} With the FiveK and HDR+ datasets, we perform experiments on two typical application scenarios: photo retouching and color/tone enhancement in camera imaging pipeline. In both applications, the target images are in the sRGB color space, have 8-bit dynamic range, and are compressed in JPG format. In the photo retouching application, the input images have the same format as the target images, while for color/tone enhancement in camera imaging pipeline, the input images are in the CIE XYZ color space \cite{wiki:CIE} and have 16-bit dynamic range without compression. It should be noted that we only learn the color/tone enhancement part of the imaging pipeline rather than learning the entire pipeline from raw data to final RGB output.

\begin{table}[t]
\footnotesize
\centering
\caption{Ablation studies on the number ($N$) of LUTs and the effect of CNN weight predictor.}
\label{table:ablation}
\begin{tabular}{|c|c|c|c|c|c|c|}
\hline
 $N$ & 1 (w/o CNN) & 1 & 2 & 3 & 4 & 5             \\\hline
PSNR &20.37 &23.15&24.86&25.21&25.26&25.29\\
SSIM &0.852 &0.884&0.917&0.922&0.924&0.926\\
$\bigtriangleup E^*$ &13.47 &9.83&8.07&7.61&7.58&7.54\\\hline
\end{tabular}
\end{table}


\textbf{Evaluation metrics.} We employ three metrics including PSNR, SSIM \cite{wang2004image} and delta E ($\bigtriangleup E^*$) to evaluate different methods. Delta E is a color difference metric defined in the CIELAB color space which is proven consistent to human perception \cite{backhaus2011color}. Contrary to PNSR and SSIM, a smaller $\bigtriangleup E^*$ means better performance.

\begin{figure}[t]
\centering
\subfigure{
\begin{minipage}[b]{0.8\linewidth}
\centering
\includegraphics[width=1.0\textwidth]{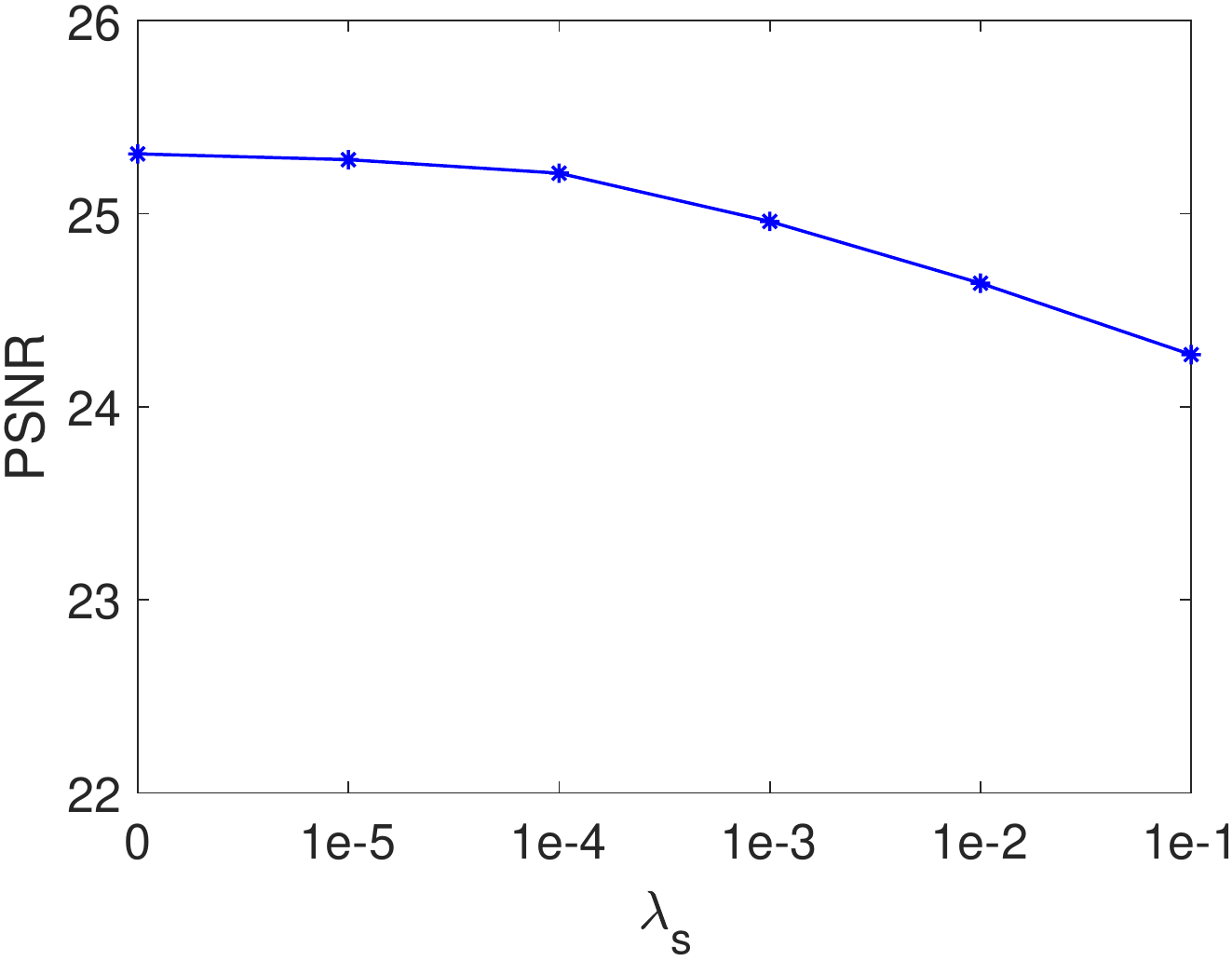}
\end{minipage}}

\subfigure{
\begin{minipage}[b]{0.8\linewidth}
\centering
\includegraphics[width=1.0\textwidth]{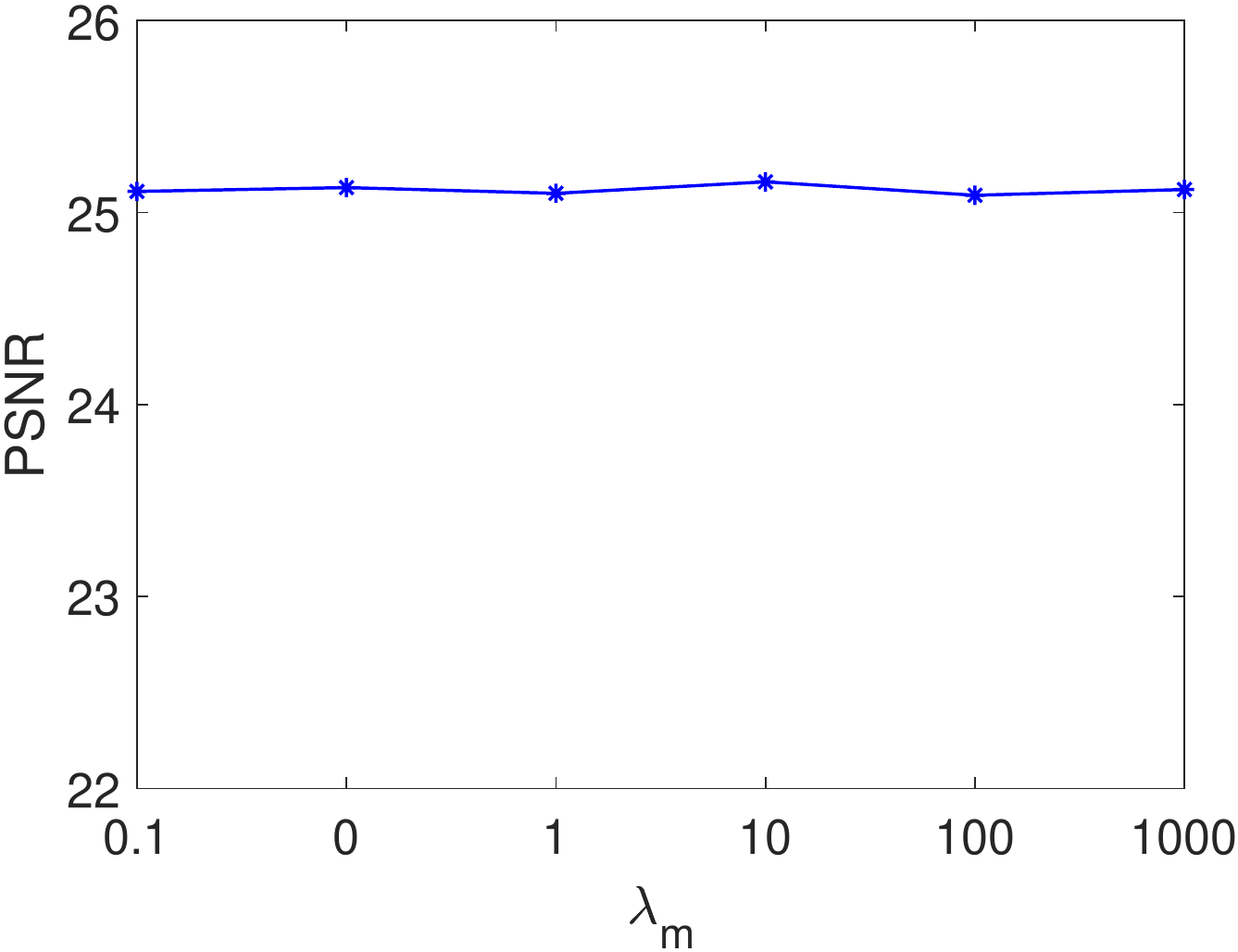}
\end{minipage}}
\caption{Effects of different choices of parameters $\lambda_s$ and $\lambda_m$.}
\label{figure:parameter_lambda}
\end{figure}

\begin{figure*}[t]
\centering
\subfigure[w/o regularization]{
\begin{minipage}[b]{0.48\linewidth}
\centering
\includegraphics[width=1.0\textwidth]{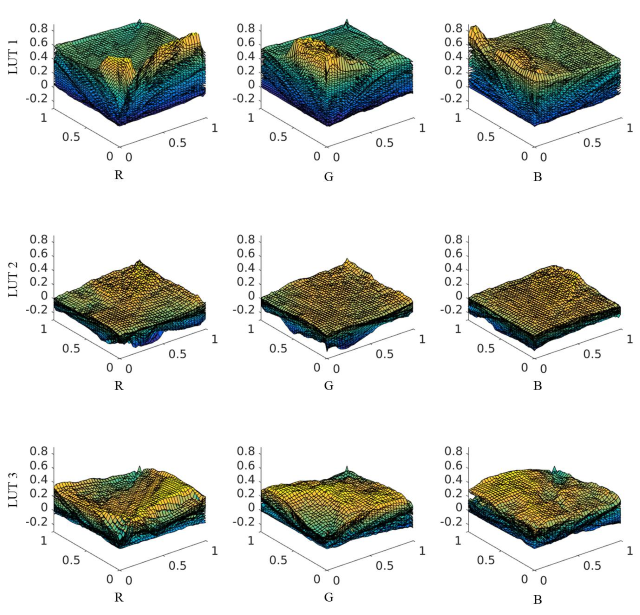}
\end{minipage}}
\subfigure[smooth only]{
\begin{minipage}[b]{0.48\linewidth}
\centering
\includegraphics[width=1.0\textwidth]{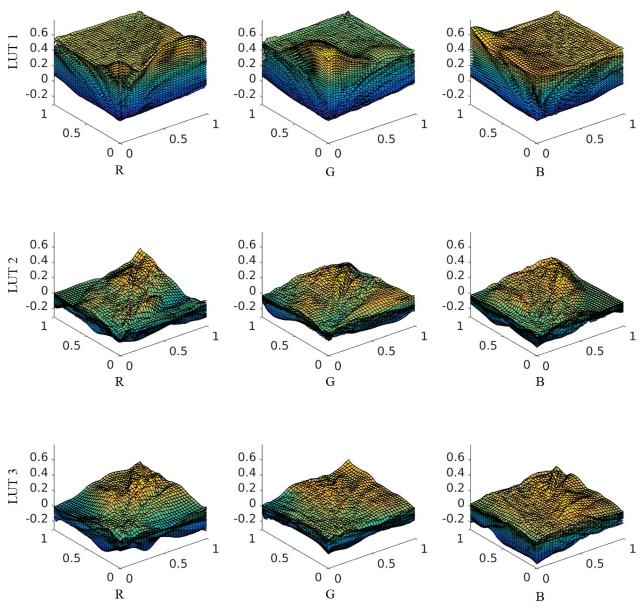}
\end{minipage}}

\vspace{-4pt}
\subfigure[monotonicity only]{
\begin{minipage}[b]{0.48\linewidth}
\centering
\includegraphics[width=1.0\textwidth]{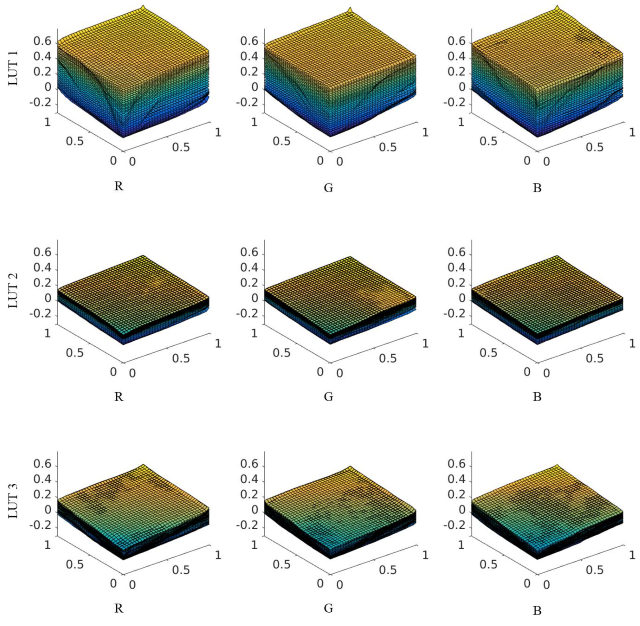}
\end{minipage}}
\subfigure[smooth + monotonicity]{
\begin{minipage}[b]{0.48\linewidth}
\centering
\includegraphics[width=1.0\textwidth]{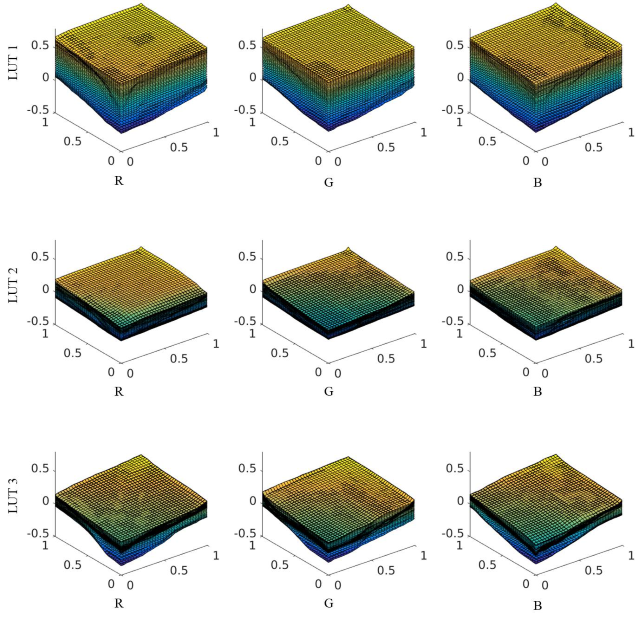}
\end{minipage}}
\vspace{-4pt}
\caption{Visualization of each of the R, G, B channels of the three basis 3D LUTs learned (a) without regularization, (b) with only smooth regularization, (c) with only monotonicity regularization and (d) with both smooth and monotonicity regularization.}
\label{figure:visualization_LUT}
\end{figure*}

\begin{figure*}[h]
\centering
\subfigure{
\begin{minipage}[b]{1.0\linewidth}
\centering
\includegraphics[width=1.0\textwidth]{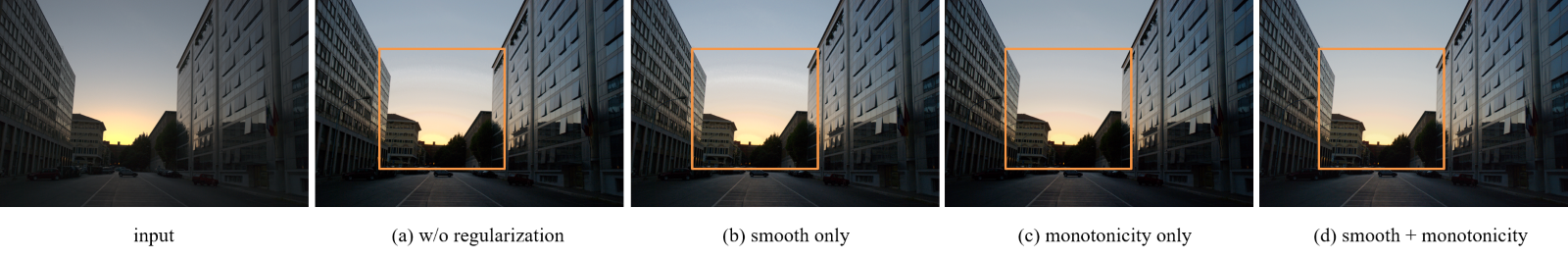}
\end{minipage}}%
\vspace{-8pt}
\caption{Enhancement results of a typical input image (a) without regularization, (b) with only smooth or (c) only monotonicity regularization, and (d) with both smooth and monotonicity regularization. Best viewed in color.}
\label{figure:regularization}
\end{figure*}

\begin{figure*}[ht]
\centering
\subfigure{
\begin{minipage}[b]{1.0\linewidth}
\centering
\includegraphics[width=1.0\textwidth]{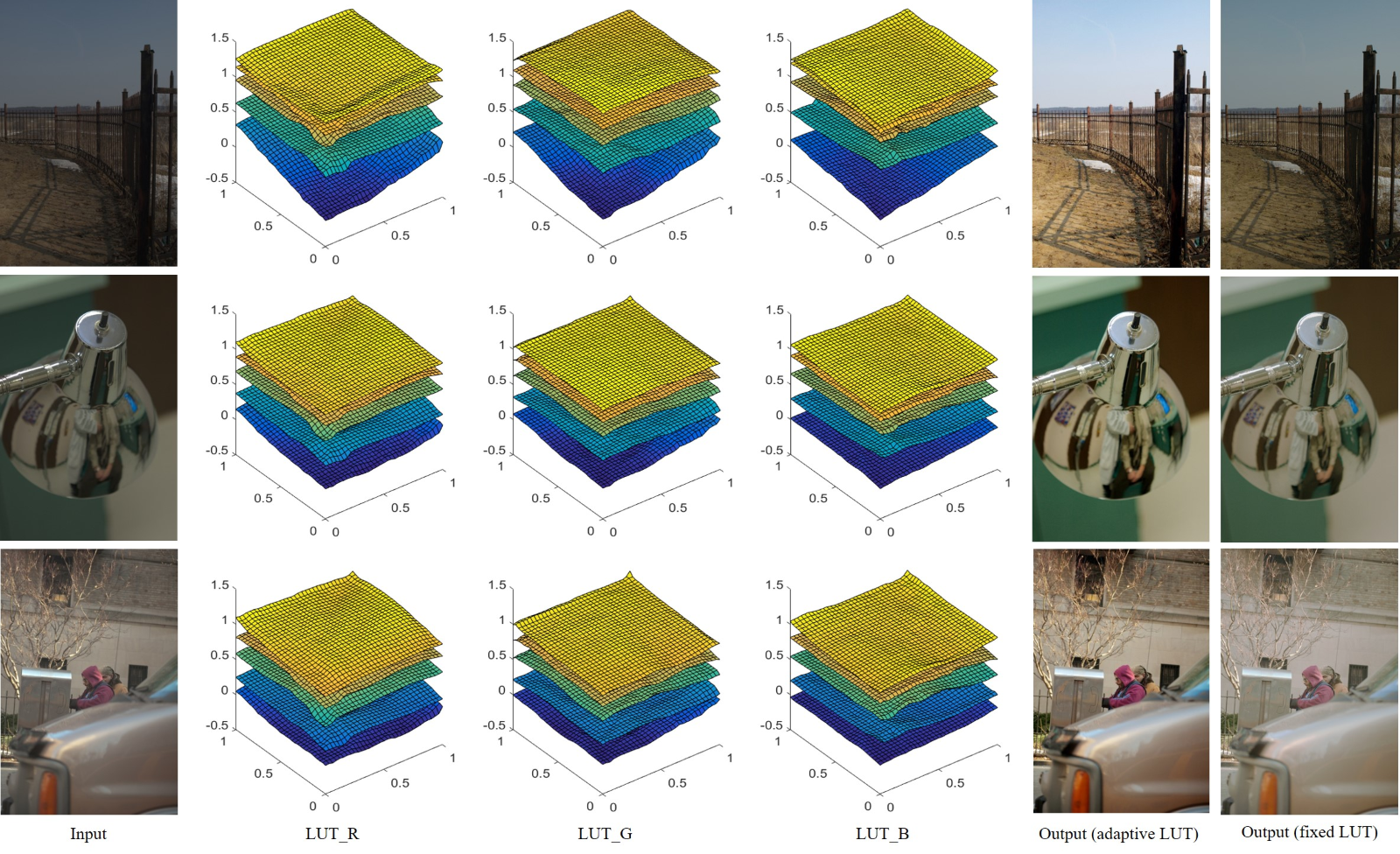}
\end{minipage}}
\vspace{-16pt}
\caption{Visualization of the image-adaptive LUTs (after combination) generated by our model for three different images and their corresponding enhancement results.}
\label{figure:visualization_adaptive_results}
\end{figure*}

\subsection{Ablation study}

We conduct ablation studies in this section to investigate the selection of the number ($N$) of 3D LUTs, analyze the roles of CNN weight predictor and regularization terms, and demonstrate the image adaptiveness of our model. We perform the ablation study experiments on the photo retouching task using paired data from the FiveK dataset, and the image resolution is set to 480p. During the ablation studies, 4,500 image pairs are used for training and 500 image pairs are used for testing. 


\textbf{The number $N$ of 3D LUTs.} To determine the number $N$ of 3D LUTs and demonstrate the role of CNN weight predictor, we evaluate the performance of our model by setting $N=1$ without CNN weight predictor, and setting $N= \{1,2,3,4,5\}$ with CNN weight predictor. The PSNR, SSIM and $\bigtriangleup E^*$ of different settings are reported in Table \ref{table:ablation}. One can see that, learning a single 3D LUT without the CNN weight predictor has poor performance (PSNR: 20.37, SSIM: 0.852 and $\bigtriangleup E^*$: 13.47) because the 3D LUT itself can only perform fixed transformation for all images. Combining a single 3D LUT with a CNN weight predictor significantly improves the performance (PSNR: 23.15, SSIM: 0.884 and $\bigtriangleup E^*$: 9.83) by making the transformation (mainly in brightness) adaptive to image content. Increasing the number of 3D LUTs from 1 to 3 noticeably boosts the performance (PSNR: 25.21, SSIM: 0.922 and $\bigtriangleup E^*$: 7.61) because using multiple 3D LUTs improves much the expressiveness of color transforms. Further increasing $N$ from 3 to 5 brings minor improvements. We thus set $N=3$ in all our following experiments for a good trade-off between performance and model compactness. Compared to using one single 3D LUT, our final model with three 3D LUTs improves the PSNR, SSIM and $\bigtriangleup E^*$ by 4.84 dB, 0.07 and 5.86, respectively, which validates the effectiveness of our design on learning image-adaptive 3D LUTs.


\textbf{Selection of regularization parameters.} In the paired and unpaired loss functions in Eqs. \ref{equ:paired} and \ref{equ:unpaired}, we introduced two constant parameters $\lambda_s$ and $\lambda_m$ to control the effects of the smooth regularization $\mathcal{R}_{s}$ and monotonicity regularization $\mathcal{R}_{m}$. Here we conduct a set of experiments to determine the suitable choices of them. Generally, the smooth regularization should be small since a heavy smoothness penalty will result in flat 3D LUTs with limited transformation flexibility, while the monotonicity regularization can be relatively stronger. We thus evaluate and select $\lambda_s$ and $\lambda_m$ from $\{0,0.00001,0.0001,0.001,0.01,0.1\}$ and $\{0.1,0,1.0,10,100,1000\}$, respectively. When evaluating $\lambda_s$, $\lambda_m$ is fixed at 10, and when evaluating $\lambda_m$, $\lambda_s$ is fixed at 0.0001. In Fig. \ref{figure:parameter_lambda}, we plot the PNSR curves w.r.t. $\lambda_s$, $\lambda_m$. We can see that a large $\lambda_s$ (e.g., $>$0.0001) leads to worse PSNR. This is easy to understand because the smooth regularization constrains the transformation flexibility of 3D LUTs. In contrast, the PSNR is insensitive to the choice of $\lambda_m$ since monotonicity is a natural constraint to 3D LUT. 


In addition to the PSNR curves, we also visualize the effect of the two regularization terms on the learned 3D LUTs and the photo enhancement results. In Fig. \ref{figure:visualization_LUT}, we visualize the three 3D LUTs learned without regularization, with only smooth regularization ($\lambda_s$=0.0001, $\lambda_m$=0), with only monotonicity regularization ($\lambda_s$=0, $\lambda_m$=10) and with both smooth and monotonicity regularization ($\lambda_s$=0.0001, $\lambda_m$=10). The corresponding enhancement results by those 3D LUTs on one typical scene are shown in Fig. \ref{figure:regularization}. As can be seen, the 3D LUTs learned without regularization (Fig. \ref{figure:visualization_LUT}(a)) have very irregular surface, which means that the output colors of these 3D LUTs may significantly change in some local areas. As a result, some visible banding artifacts appear in the transitional sky area in Fig. \ref{figure:regularization}(a). Employing only the smooth regularization results in locally smoother, however still irregular, 3D LUTs because some parts of the color space are not activated by the training data. Employing only the monotonicity regularization leads to globally more regular 3D LUTs and alleviates much the banding artifacts; however, some local areas are not smooth enough. Using both the regularization terms lead to smoother and much more regular 3D LUTs, which effectively suppresses the banding artifacts, as shown in Fig. \ref{figure:regularization}(d).


Based on the above quantitative and qualitative results, we choose $\lambda_s$=0.0001 and $\lambda_m$=10 in our following experiments, which deliver a good trade-off between enhancement performance and model stability.

\textbf{Image-adaptive property.} As we described in Section \ref{section:image-adaptive}, our method is able to generate an adaptive 3D LUT (by fusing adaptively the three basis LUTs) to the input image, and uses it to enhance the given image. To better study this image-adaptive enhancement property, in Fig. \ref{figure:visualization_adaptive_results} we visualize the image-adaptive 3D LUTs for three different images and show their corresponding enhancement results. In order to more clearly see the difference on the adaptive LUTs, we only visualize five slices (i.e., $\{1, 9, 17, 25, 33\}$) of the whole 3D LUT (33 slices in total). As a reference, we also show the enhancement results obtained by learning one fixed 3D LUT (without CNN) using the same training data on the right column. As can be seen, the adaptive 3D LUT varies with the different image content to better enhance their brightness, color and tone. In contrast, the fixed 3D LUT enhances different images in the same manner without considering their different content, resulting in less satisfactory brightness, hue and saturation. 


\subsection{Experiment on photo retouching}

\begin{figure*}[ht]
\centering
\subfigure{
\begin{minipage}[b]{1.0\linewidth}
\centering
\includegraphics[width=1.0\textwidth]{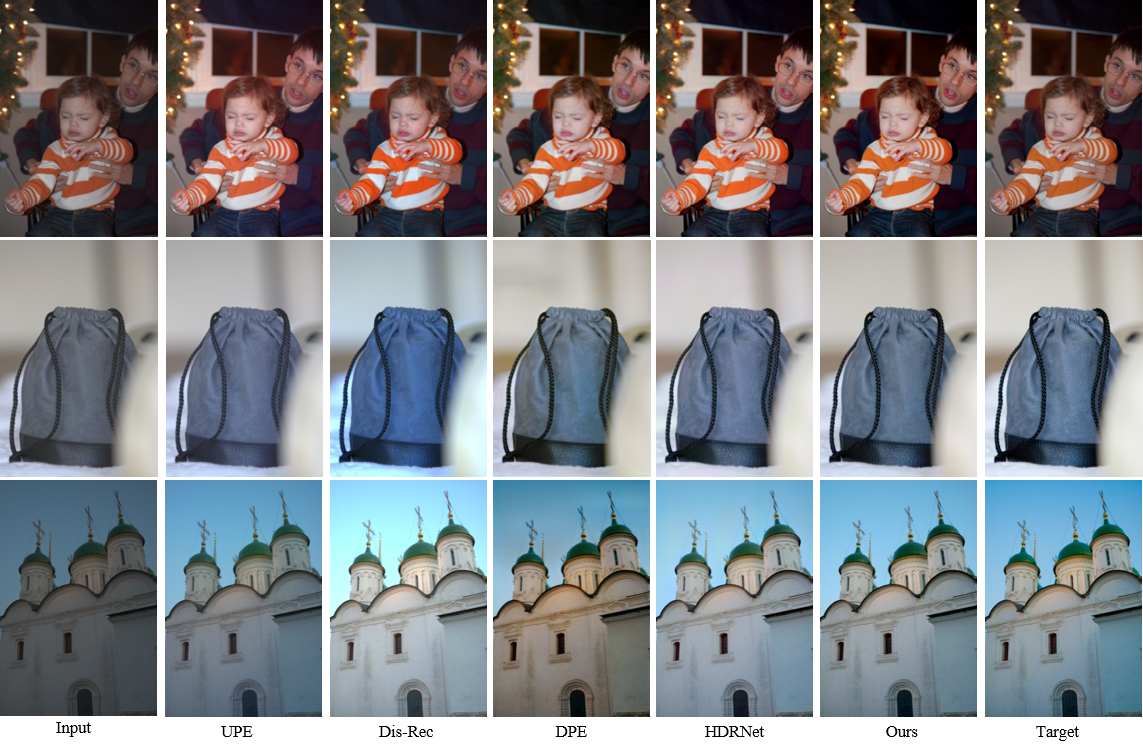}
\end{minipage}}
\vspace{-16pt}
\caption{Qualitative comparison of different paired learning methods for photo retouching on the FiveK dataset.}
\label{figure:photo_retouching_paired}
\end{figure*}

\begin{figure*}[h]
\centering
\subfigure{
\begin{minipage}[b]{1.0\linewidth}
\centering
\includegraphics[width=1.0\textwidth]{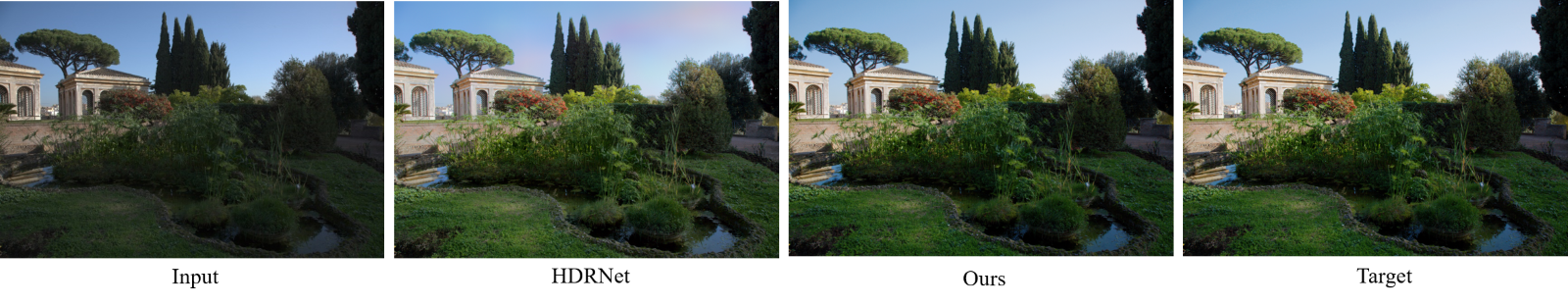}
\end{minipage}}
\vspace{-16pt}
\caption{Qualitative comparison between HDRNet and our method for photo retouching on the FiveK dataset. Note that some sky areas enhanced by HDRNet are biased to purple.}
\label{figure:HDRNet_ours}
\end{figure*}

This section compares our method with the latest deep enhancement methods for photo retouching on the FiveK dataset. We present the results under the pairwise learning and unpaired learning settings. Among previous methods, only the work in \cite{chen2018deep} considers both paired and unpaired learning. We thus use the same split as in \cite{chen2018deep} in our experiments for fair comparison. Specifically, in pairwise learning, 4,500 image pairs are used for training and the remaining 500 image pairs are used for testing; in unpaired learning, 2,250 images are used as the source data and 2,250 retouched images (from other source images with different content) are used as the target data. The remaining 500 images are used for testing. For all competing methods, we used the source code released by the authors and retrained their models using the same data split and format. We trained each model until it is converged and report their best results on the testing set. 


\textbf{Learning with paired data.} Under the pairwise learning setting, we compare our method with the latest pairwise photo enhancement methods whose source codes are available: HDRNet \cite{gharbi2017deep}, DPE \cite{chen2018deep}, Dis-Rec \cite{park2018distort} and UPE \cite{Wang2019CVPR}. HDRNet learns pixel-wise transformation coefficients in the bilateral space, DPE learns pixel-to-pixel mapping using a U-Net architecture, Dis-Rec employs the reinforcement learning to perform iterative enhancement and UPE estimates pixel-wise illumination using the same architecture of HDRNet.

As mentioned in Section \ref{subsection:experimental setup}, we train and evaluate each method on two resolutions: 480p and the image original resolution (with about 12 megapixels). We randomly cropped $256\times256$ and $1024\times1024$ patches on 480p and original resolution, respectively, to train the competing models. The PSNR, SSIM and $\bigtriangleup E^*$ indexes are reported in Table \ref{table:quantitative pairwise retouching}. The results of DPE on original resolution are unavailable because the DPE method is too memory demanding to implement using our Titan RTX GPU with 24 GB memory on this resolution. As can been seen from Table \ref{table:quantitative pairwise retouching}, our method significantly outperforms all the competing methods on both resolutions in terms of all the three metrics. Specifically, our method has an advantage of $\geq$0.9 dB in PSNR than the second best method HDRNet \cite{gharbi2017deep} on 480p. The advantage becomes higher (1.1dB) when working on the original image resolution, indicating the robustness of using 3D LUT to process high-resolution photos. DPE can also obtain reasonable performance on the 480p resolution. However, the heavy computational cost of U-Net architecture severely constrains its application on processing high-resolution images. Both Dis-Rec and UPE obtain less satisfactory performance, probably because of the unstability of reinforcement learning and inaccurate illumination estimation, respectively.


Fig. \ref{figure:photo_retouching_paired} presents qualitative comparisons of different paired learning methods. One can have the following observations. UPE \cite{chen2018deep} only moderately enhances the input image, generating outputs with insufficient saturation and contrast. Dis-Rec \cite{park2018distort} is not robust enough to handle different scenes, resulting in over-exposed areas or unnatural color cast. DPE \cite{chen2018deep} may generate halo artifacts or unnatural color cast. Both HDRNet \cite{gharbi2017deep} and our method are able to more robustly enhance images in different scenes. A more detailed comparison between HDRNet and our method on an image is shown in Fig. \ref{figure:HDRNet_ours}. It can be seen that some sky areas enhanced by HDRNet are biased to purple. One possible reason is that the HDRNet model overfits the training data where most sky content are either blue or red. As a consequence, the HDRNet enhances more on the R and B channels than the G channel in the sky area, resulting in purple (R and B are significantly larger than G) color cast. In contrast, our model does not overfit the training data, having better generalization performance than HDRNet. This owes to the compact model design of our method as well as the smooth and monotonicity regularizations in model training.


\begin{table}[t]
\footnotesize
\centering
\caption{Quantitative comparison of photo retouching results under the pairwise training setting on FiveK dataset. ``N.A." means that the result is not available.}
\label{table:quantitative pairwise retouching}
\begin{tabular}{|c|ccc|ccc|}
\hline
\multirow{2}{*}{Method} & \multicolumn{3}{c|}{FiveK (480p)} & \multicolumn{3}{c|}{FiveK (original)}         \\\cline{2-7}
               & PSNR & SSIM & $\bigtriangleup E^*$   & PSNR    & SSIM &$\bigtriangleup E^*$        \\\hline
UPE \cite{Wang2019CVPR} & 21.88 & 0.853 & 10.80 & 21.65 & 0.859 & 11.09\\
Dis-Rec \cite{park2018distort} & 21.98 & 0.856& 10.42  &21.81&0.862 & 10.60\\
DPE \cite{chen2018deep} & 23.75 & 0.908 & 9.34 & N.A. & N.A. & N.A.\\
HDRNet \cite{gharbi2017deep} & 24.32 & 0.912  & 8.49 & 24.03 & 0.919 & 8.68\\\hline\hline
Ours & 25.21 & 0.922 & 7.61  & 25.10 & 0.930 & 7.72\\\hline
\end{tabular}
\end{table}

\begin{table}[t]
\footnotesize
\centering
\caption{Quantitative comparison of photo retouching results under the unpaired training setting on the FiveK dataset. ``N.A." means that the result is not available.}
\label{table:quantitative unpaired retouching}
\begin{tabular}{|c|ccc|ccc|}
\hline
\multirow{2}{*}{Method} & \multicolumn{3}{c|}{FiveK (480p)} & \multicolumn{3}{c|}{FiveK (original)}          \\\cline{2-7}
& PSNR & SSIM & $\bigtriangleup E^*$    & PSNR    & SSIM  &$\bigtriangleup E^*$               \\\hline
Camera Raw & 21.61 & 0.854 & 11.83 & 21.55 & 0.861 & 11.98 \\
Pix2Pix \cite{isola2017image} & 19.21 & 0.814 & 14.76& N.A. & N.A. & N.A. \\
CycGAN \cite{zhu2017unpaired} & 20.98 & 0.831 & 13.28& N.A. & N.A.& N.A. \\
White-Box \cite{hu2018exposure} & 21.32 & 0.864 & 12.65& 21.17 & 0.875 &12.81\\
DPE \cite{chen2018deep} & 21.99 & 0.875 & 11.40 & N.A. & N.A.& N.A. \\
UIE \cite{kosugi2020unpaired} & 22.11& 0.879 & 11.21 & 22.03& 0.882 &11.46\\
\hline\hline
Ours & 22.86 & 0.887 & 10.28 & 22.78 & 0.898 & 10.42\\\hline
\end{tabular}
\end{table}

\begin{figure*}[ht]
\centering
\subfigure{
\begin{minipage}[b]{1.0\linewidth}
\centering
\includegraphics[width=1.0\textwidth]{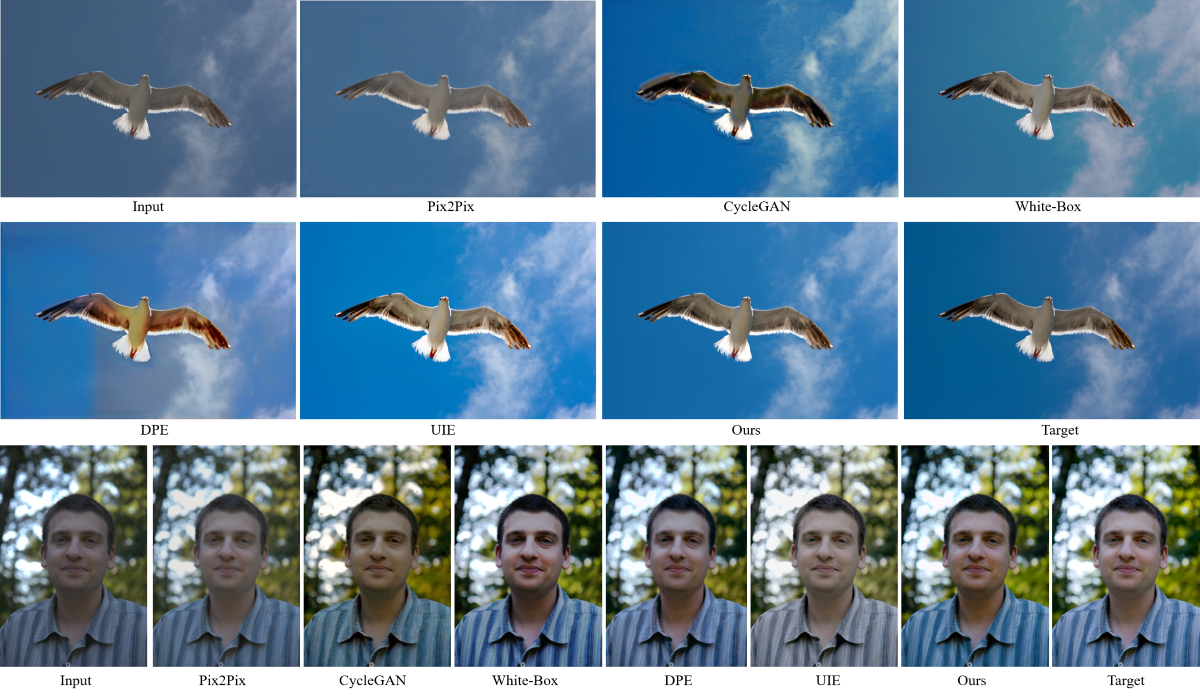}
\end{minipage}}
\caption{Qualitative comparison of different unpaired learning methods for photo retouching on the FiveK dataset.}
\label{figure:photo_retouching_unpaired}
\end{figure*}

\textbf{Learning with unpaired data.} Under the unpaired learning setting, we compare our method with Pix2Pix \cite{isola2017image}, CycleGAN \cite{zhu2017unpaired}, DPE \cite{chen2018deep}, White-Box \cite{hu2018exposure} and UIE \cite{kosugi2020unpaired}. Pix2Pix and CycleGAN are two representative image-to-image translation methods, DPE learns pixel-to-pixel transformation with a U-Net architecture carefully tuned for photo enhancement, both White-Box and UIE learn to enhance input images using a set of basic operations in an iterative manner. We also report the results obtained by the automatic adjustment of Camera Raw in Photoshop as a reference. All models are retrained and evaluated using the same training and testing splits. All parameters (e.g., CNN model, 3D LUTs, regularization terms) of our method were kept the same as those in paired learning.

The quantitative results of all competing methods are reported in Table \ref{table:quantitative unpaired retouching}. Note that the results of Pix2Pix, CycleGAN and DPE on the original resolution are unavailable because of the out of memory problem. Again, our method achieves much better PSNR, SSIM and $\bigtriangleup E^*$ indices on both resolutions than the other methods. Specifically, our model outperforms the second best one UIE by 0.75 dB, 0.016 and 1.04 on PSNR, SSIM and $\bigtriangleup E^*$, respectively. An interesting observation is that on these objective metrics, most of the unpaired enhancement methods, except ours, do not have obvious advantage over the automatic adjustment of Camera Raw which does not leverage any training data of FiveK dataset. The main reason is that most unpaired methods are not stable enough, resulting in low indices on some failure cases. Fig. \ref{figure:photo_retouching_unpaired} presents the visual comparisons of competing methods on two typical scenes. One can see that the Pix2Pix \cite{isola2017image} only slightly enhances the input image. The CycleGAN \cite{zhu2017unpaired}, White-Box \cite{hu2018exposure}, DPE \cite{Wang2019CVPR} and UIE \cite{kosugi2020unpaired} are not stable in different scenes. They may result in over- or under-enhanced images, produce unnatural color or distorted image content. In contrast, our method can stably and properly enhance images under different scenes. 


\subsection{Experiment on imaging pipeline enhancement}

\begin{table}[t]
\footnotesize
\centering
\caption{Quantitative comparison of different pairwise learning methods for imaging pipeline enhancement on the FiveK and HDR+ datasets.}
\label{table:quantitative pairwise pipeline}
\begin{tabular}{|c|ccc|ccc|}
\hline
\multirow{2}{*}{Method} & \multicolumn{3}{c|}{FiveK (480p)} & \multicolumn{3}{c|}{HDR+ (480p)}               \\\cline{2-7}
& PSNR & SSIM &$\bigtriangleup E^*$ & PSNR    & SSIM &$\bigtriangleup E^*$                 \\\hline
UPE \cite{Wang2019CVPR} & 21.56 & 0.837 & 12.29 & 21.21 & 0.816 & 13.05\\
DPE \cite{chen2018deep} & 22.93 & 0.894 & 11.09 & 22.56 & 0.872 & 10.45\\
HDRNet \cite{gharbi2017deep} &24.14 &0.913& 8.65& 23.04 & 0.879 & 8.97\\\hline\hline
Ours & 25.06 & 0.920& 7.63 & 23.54 & 0.885 & 7.93\\\hline
\end{tabular}
\end{table}

\begin{table}[t]
\footnotesize
\centering
\caption{Quantitative comparison of different unpaired learning methods for imaging pipeline enhancement on the FiveK and HDR+ datasets.}
\label{table:quantitative unpaired pipeline}
\begin{tabular}{|c|ccc|ccc|}
\hline
\multirow{2}{*}{Method} & \multicolumn{3}{c|}{FiveK (480p)} & \multicolumn{3}{c|}{HDR+ (480p)}               \\\cline{2-7}
& PSNR & SSIM &$\bigtriangleup E^*$& PSNR    & SSIM &$\bigtriangleup E^*$    \\\hline
Camera Raw & 21.61 & 0.854 & 11.83 & 19.86 & 0.791 & 14.98 \\
Pix2Pix \cite{isola2017image} & 14.91 & 0.678 &27.26& 13.77 & 0.593& 30.94\\
CycGAN \cite{zhu2017unpaired} & 17.01 & 0.729 &23.64 & 15.61 & 0.647& 28.15\\
White-Box \cite{hu2018exposure} & 17.62 & 0.752 &20.47& 16.47 & 0.689& 26.12\\
DPE \cite{chen2018deep} & 17.68 & 0.764 &20.13& 16.56 & 0.698 & 25.56\\
UIE \cite{kosugi2020unpaired} & 17.83 & 0.772 & 19.74 & 16.72 & 0.706  & 25.02\\\hline\hline
Ours & 21.60 & 0.852 &11.95 & 18.98 & 0.767& 16.21 \\\hline
\end{tabular}
\end{table}

\begin{figure*}[ht]
\centering
\subfigure{
\begin{minipage}[b]{1.0\linewidth}
\centering
\includegraphics[width=1.0\textwidth]{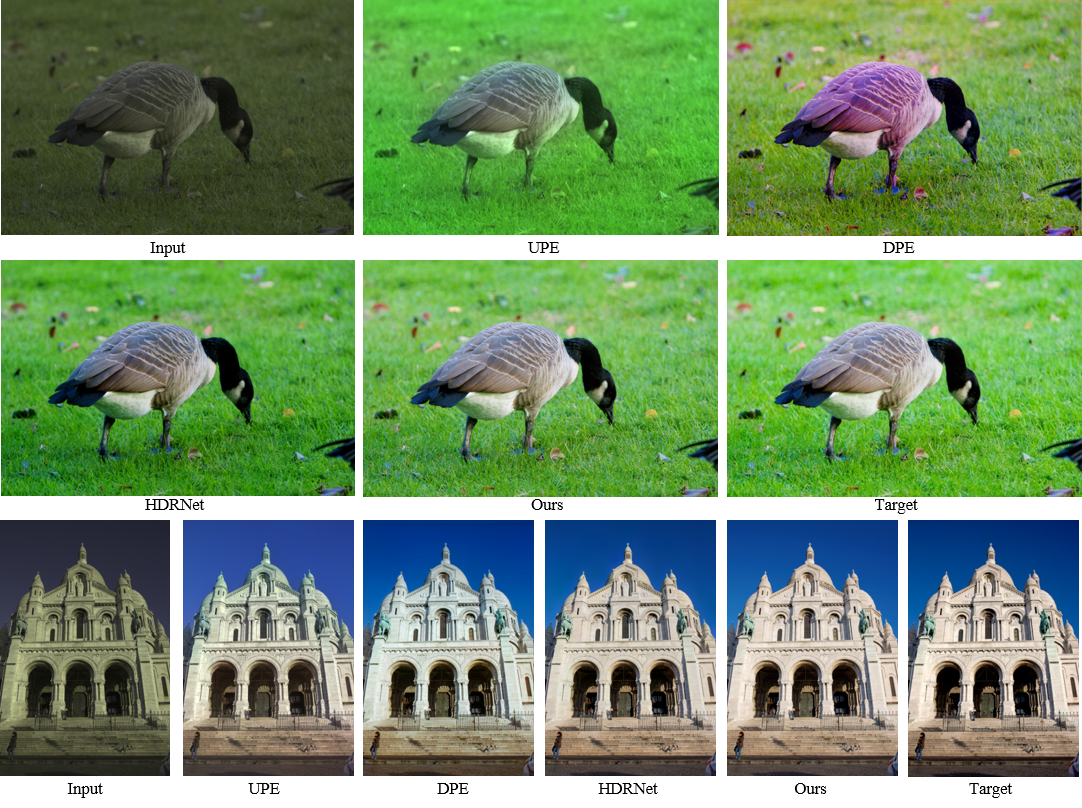}
\end{minipage}}
\caption{Qualitative comparison of different paired learning methods for imaging pipeline enhancement on the FiveK and HDR+ datasets.}
\label{figure:pipeline_enhancement_paired}
\end{figure*}

In this section we compare our method with the latest deep enhancement methods for color/tone enhancement in camera imaging pipeline. In this case, the input images are 16-bit uncompressed images in the CIE XYZ color space. For pairwise learning, all the settings are the same as those used in the photo retouching task. For unpaired learning, we change $\lambda_1$ in Eq. \ref{equ:generator} to 10 for all competing methods (using a GAN loss) to relax the content preservation constraint since the input images are very different from the target images in this set of experiments (e.g., different dynamic range, different color space, uncompressed vs. compressed). All competing models are retrained on this task. 

We conduct experiments on the FiveK and HDR+ datasets. On the FiveK dataset, data splits are kept the same as those in photo retorching task. Since most scenes in the HDR+ dataset are not aligned between the intermediate frame and the groundtruth, we conduct experiments using the well aligned Nexus 6p subset, among which 675 scenes are used for training and 250 scenes are used for testing in pairwise setting. In the unpaired setting, we randomly divide the 675 scenes into a source domain and a target domain without overlapped content. Intermediate frame from the source domain and groundtruth from the target domain are used to train unpaired models. To compare with as many methods as possible, images are resized to 480p resolution.


\textbf{Learning with paired data.} We compare our model with HDRNet \cite{gharbi2017deep}, DPE \cite{chen2018deep}, and UPE \cite{Wang2019CVPR} under the pairwise learning setting. The Dis-Rec \cite{park2018distort} is not included in this comparison since it is difficult to modify its source code to process 16-bit input images in CIE XYZ color space. The results on the FiveK and HDR+ datasets are reported in Table \ref{table:quantitative pairwise pipeline}. On the FiveK dataset, the performance on all the three metrics decreases slightly compared to the photo retouching task since imaging pipeline enhancement is more challenging. Nonetheless, our method again has clear advantage over all competing methods on both datasets. For all methods, the performance on the HDR+ dataset is much lower than on the FiveK dataset owning to two major reasons. First, the number of training image pairs is much less on the HDR+ dataset. Second, some images captured by mobile cameras in low-light environment contain heavy noise while all the enhancement methods do not have denoising function. 

In Fig. \ref{figure:pipeline_enhancement_paired}, we show the visual results obtained by different methods on two scenes. Similar conclusion can be drawn as in the photo retouching task: both our method and the HDRNet can stably enhance different inputs while the HDRNet may result in over-saturated color cast (see the greensward) because of its over-fitting. The enhancement results of our model are more nature and closer to the targets. In contrast, both UPE and DPE are not stable and may dramatically change the color.

\textbf{Learning with unpaired data.} The results of learning with unpaired data on the FiveK and HDR+ datasets are reported in Table \ref{table:quantitative unpaired pipeline}. One can see that all the competing methods obtain worse performance than the results obtained by the automatic mode of Camera Raw in this set of experiments, which may be caused by the following reason. This unpaired learning task of imaging pipeline enhancement is too challenging due to the significant differences in dynamic range and color space between input and target data. The content preservation constraint in the GAN loss (refer to Eq. \ref{equ:generator}) can hardly hold in such case. However, relaxing the content preservation constraint will enlarge much the transformation space and hurt the stability of the GAN loss. Consequently, most of the competing methods using very deep neural networks and/or complicated reinforcement learning strategy fail to learn a stable transformation. In contrast, with effective regularization terms, our model learns compact 3D LUTs and a small CNN, which still achieves reasonable performance on the FiveK dataset (almost the same as camera raw), outperforming the competing methods by a large margin ($>$ 4 dB). This further validates the stability and generalization capability of our method. The performance of all methods drops much on the HDR+ dataset because of the insufficient training data.

\subsection{User study}

\begin{figure*}[t]
\centering
\subfigure[]{
\begin{minipage}[b]{0.36\linewidth}
\centering
\includegraphics[width=1.0\textwidth]{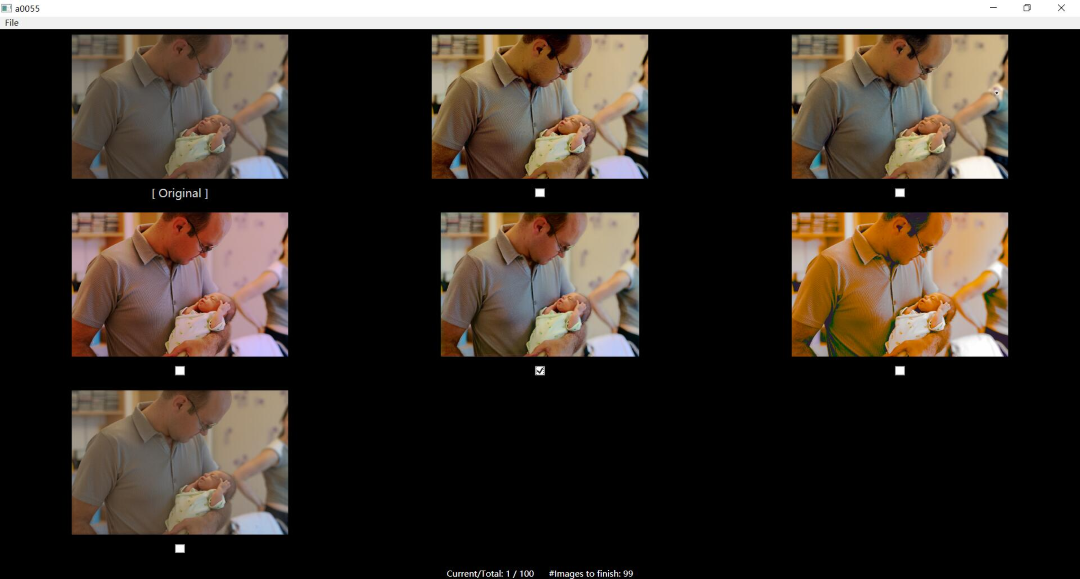}
\end{minipage}}
\subfigure[]{
\begin{minipage}[b]{0.25\linewidth}
\centering
\includegraphics[width=1.0\textwidth]{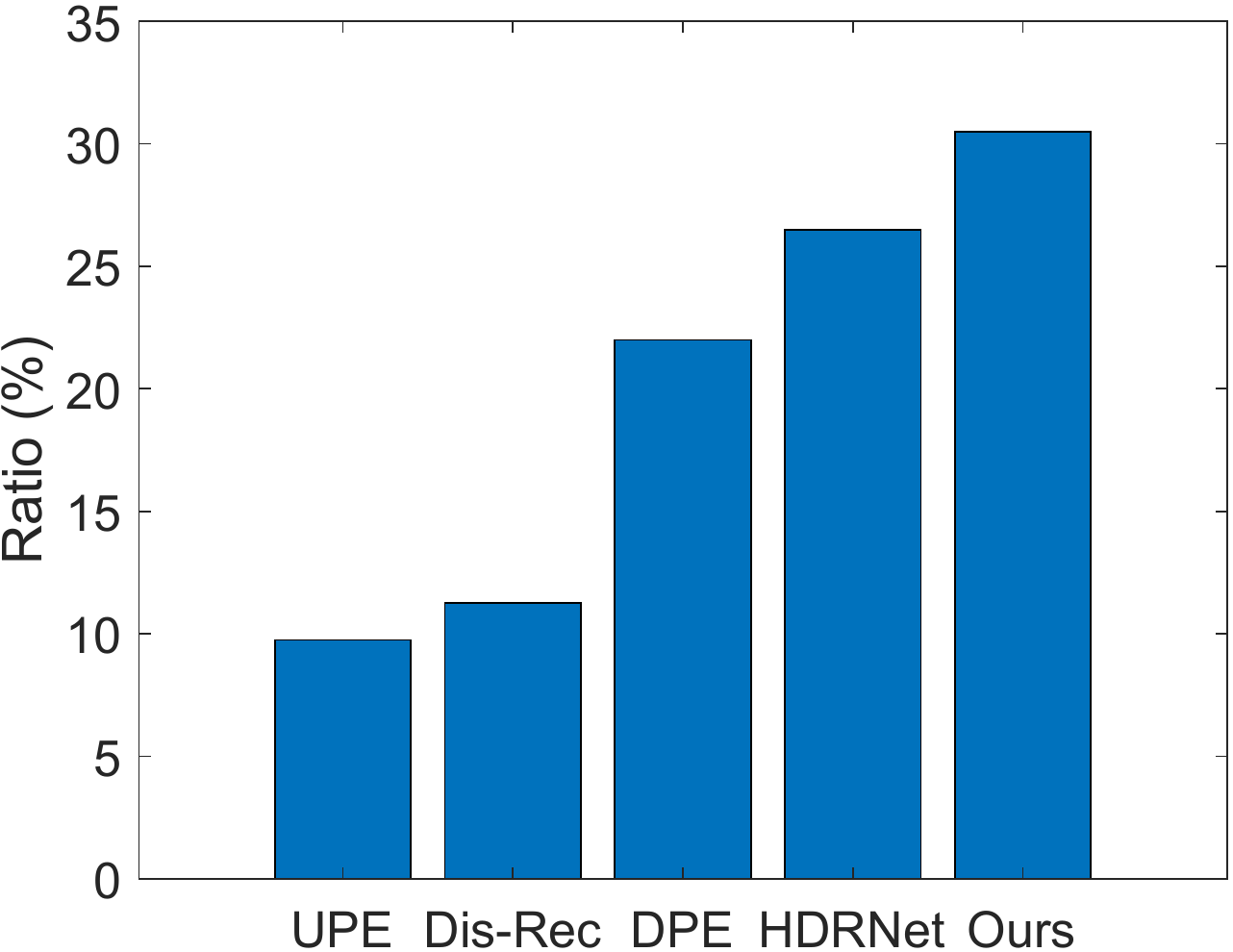}
\end{minipage}}
\subfigure[]{
\begin{minipage}[b]{0.36\linewidth}
\centering
\includegraphics[width=1.0\textwidth]{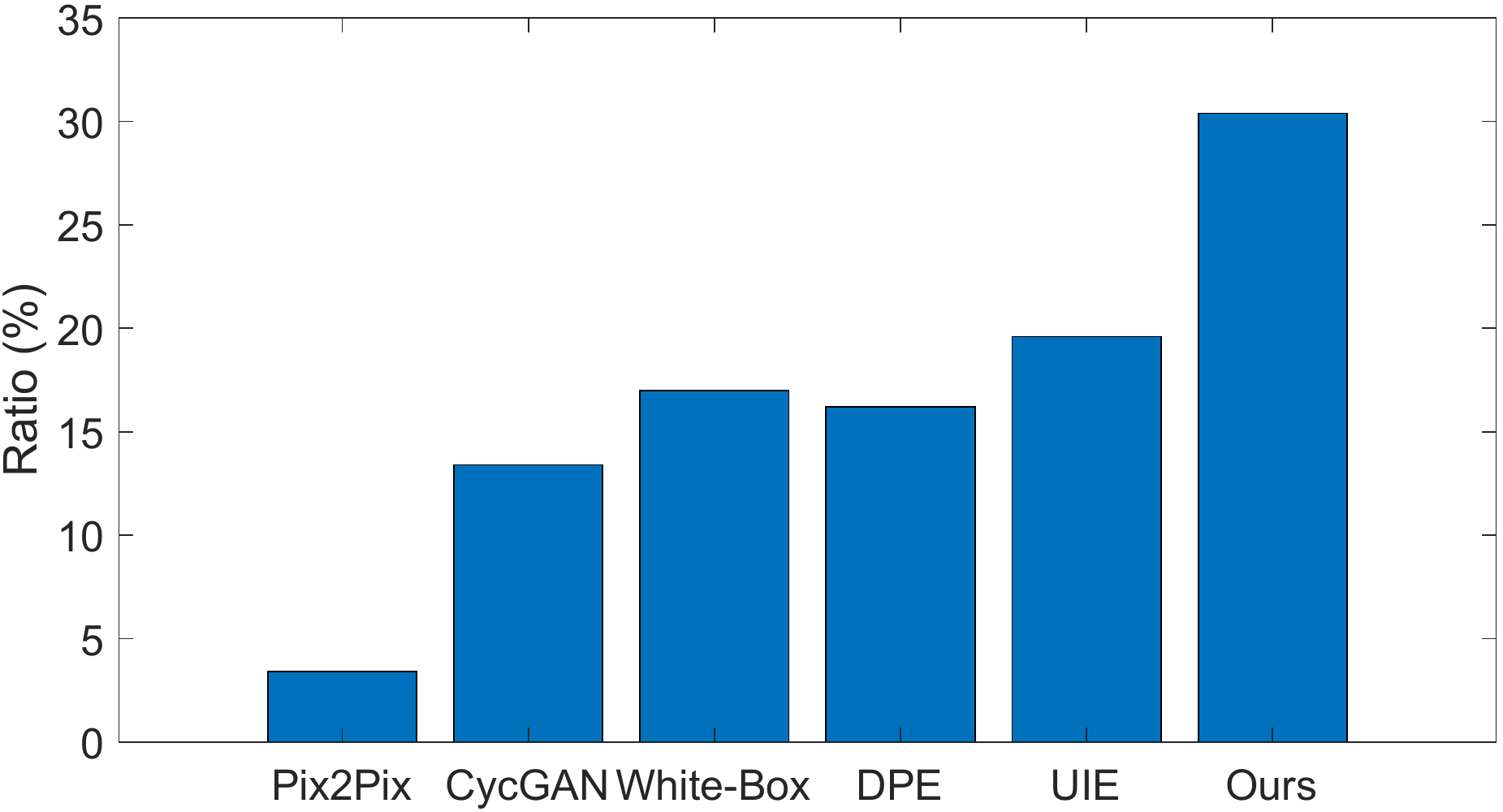}
\end{minipage}}
\caption{(a) Interface of the user study tool; (b) voting statistics of paired learning methods; and (c) voting statistics of unpaired  learning methods.}
\label{figure:user_study}
\end{figure*}

We also conduct two sets of user studies to evaluate more comprehensively the subjective perception of enhancement results obtained by different models on the paired and unpaired photo retouching tasks. To enable these studies, we developed a user study tool whose interface is shown in Fig.\ref{figure:user_study}(a). This tool supports comparison of at most 9 images (i.e., a testing image and the output images of at most 8 methods) at each time. We randomly selected 100 images from the 500 testing images of FiveK dataset and invited 20 human subjects (including postgraduate students and undergraduate students, 11 males and 9 females) to participate in our user study. For each testing image, the enhancement results obtained by different models are presented in random order and the participants are asked to select the best (visually appealing and natural) one among all competitors. 
We summarize the voting results of all participants and plot the normalized voting ratios on paired and unpaired photo retouching in Figs. \ref{figure:user_study}(b) and (c), respectively. 

\begin{figure*}[ht]
\centering
\subfigure{
\begin{minipage}[b]{1.0\linewidth}
\centering
\includegraphics[width=1.0\textwidth]{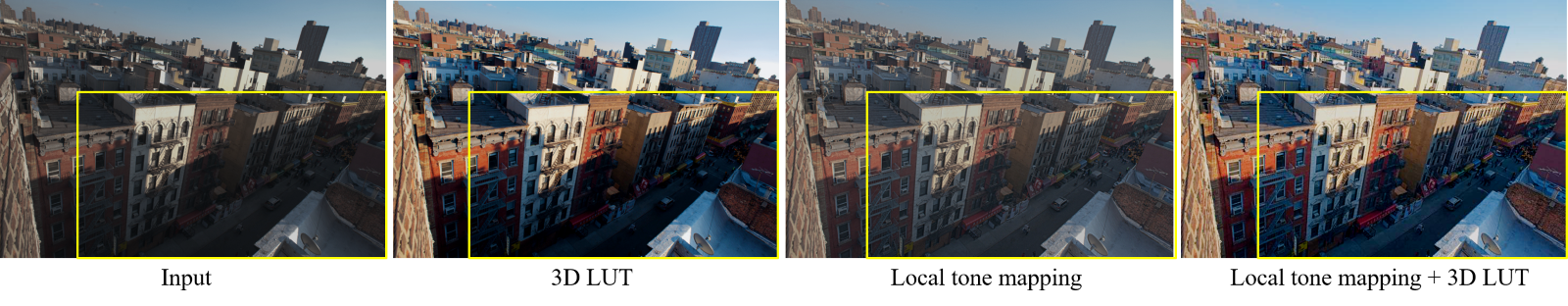}
\end{minipage}}
\vspace{-16pt}
\caption{Enhancement results on an input scene with very high dynamic range. Our 3D LUT model effectively enhances the global color and tone, while the local contrast in the shadow area can be further improved by combining some local tone mapping operator with our model.}
\label{figure:limitation_localcontrast}
\end{figure*}

\begin{figure*}[ht]
\centering
\subfigure{
\begin{minipage}[b]{1.0\linewidth}
\centering
\includegraphics[width=1.0\textwidth]{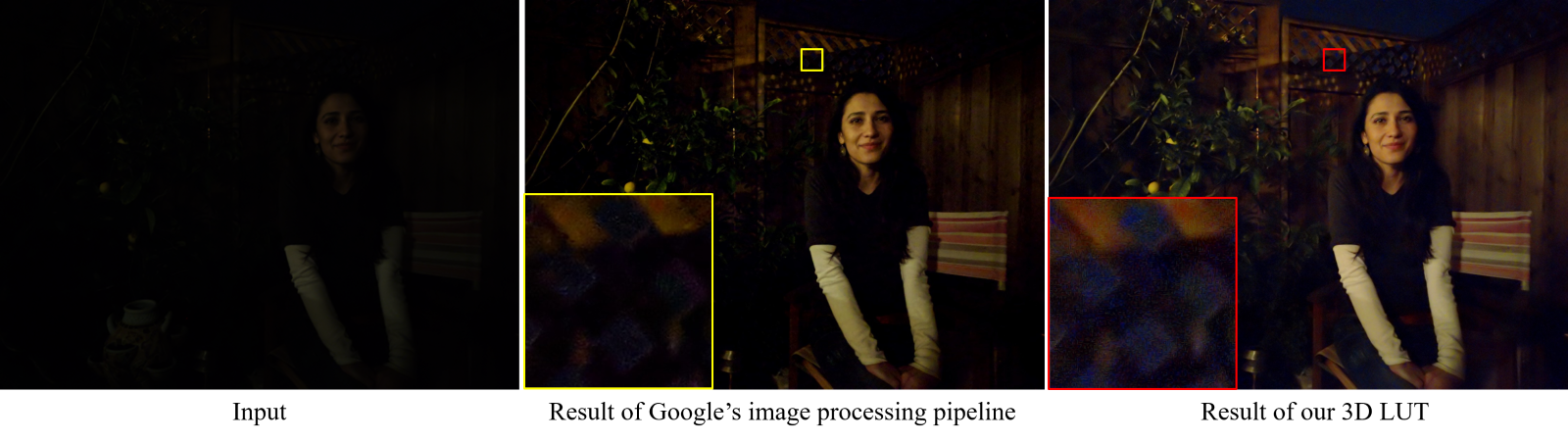}
\end{minipage}}
\vspace{-16pt}
\caption{Enhancement results generated by our 3D LUT model and Google's image processing pipeline on a noisy input (with 16-bit dynamic range) from the HDR+ dataset. Although our model properly enhances the dark area, the noise is also amplified after the enhancement. Please zoom in to see the details.}
\label{figure:limitation_noise}
\end{figure*}

Under the pairwise setting, one can see that the subjective results in  Fig. \ref{figure:user_study}(b) are basically consistent with the quantitative results in Table \ref{table:quantitative pairwise retouching}. Our method obtains the highest ratio of $30.5\%$, followed by HDRNet with $26.6\%$ and DPE with $22.1\%$. UPE and Dis-Rec receive relatively lower voting ratio (about $10\%$). Considering that the human's preference on the color and tone of photos are highly subjective and models trained with paired datasets are relatively stable in most cases, an advantage of about $4\%$ is a noticeable improvement under the paired learning setting.
Our model has more significant advantage over the competing methods in the unpaired learning setting as shown in Fig. \ref{figure:user_study}(c). Specifically, our model receives $30.4\%$ votes which is much higher than the $19.6\%$ votes obtained by the second best method UIE. Such an advantage mainly owns to the high stability of our model under the unpaired learning setting while most of the other methods generate noticeable artifacts or unnatural color cast. 

\subsection{Running speed}

Apart from the superior enhancement quality to state-of-the-art photo enhancement models, another important advantage of our method lies in its high efficiency. In this section we evaluate the running speed of competing methods. The test images are resized to three standard resolutions: 1920$\times$1080, 3840$\times$2160 and 6000$\times$4000. All models are tested on the same image set for 100 times using one Titan RTX GPU with 24 GB memory. The average running time of all methods measured by milliseconds (ms) are reported in Table \ref{table:running speed}. As can be seen, the speed of our method is at least two-orders faster than all competing methods when testing on the same GPU device and the speed advantage of our method consistently increases with the increase of image resolution. In particular, our model spends less than 2 ms in processing an image with 4K resolution, which satisfies the real-time requirement of practical applications. It is worth mentioning that HDRNet can be significantly accelerated with customized OpenGL shaders.

The high efficiency of our method mainly comes from the facts that the computational cost of the CNN weight predictor is fixed in our model and its memory cost is much lower than all competing methods. Although the CNN models in HDRNet, White-Box, Dis-Rec and UIE also have fixed input size, the HDRNet consumes intermediate computation and memory cost to calculate pixel-wise transformation coefficients on the high-resolution input image, and  White-Box, Dis-Rec and UIE need many iterations to generate the enhanced image. In contrast, our model directly maps input RGB values to output RGB values in one single forward step, and the transformation via 3D LUTs can be highly parallel using GPU.

\begin{table}[t]
\footnotesize
\centering
\caption{Running time (in milliseconds) comparison between our model and state-of-the-art deep enhancement models at different resolutions. All models are tested using one Titan RTX GPU. ``N.A." means that the result is not available.}
\label{table:running speed}
\begin{tabular}{|c|c|c|c|}
\hline
Resolution & 1920$\times$1080 & 3840$\times$2160 & 6000$\times$4000             \\\hline
Pix2Pix \cite{isola2017image} & 1.2e2 & N.A. & N.A.  \\
CycGAN \cite{zhu2017unpaired} & 5.6e2 & N.A. & N.A.  \\
DPE \cite{chen2018deep}  & 8.6e1 & N.A. & N.A. \\
White-Box \cite{hu2018exposure}  & 5.0e3 & 9.1e3 & 2.0e4 \\
Dis-Rec \cite{park2018distort}  & 2.5e4 & 1.1e5 & 3.3e5 \\
UIE \cite{kosugi2020unpaired}   & 1.0e4  & 2.0e4  & 3.3e4 \\
HDRNet \cite{gharbi2017deep}  & 4.5e1 & 2.1e2 & 5.9e2 \\
UPE \cite{Wang2019CVPR}  & 4.5e1 & 2.1e2 & 5.9e2\\\hline\hline
Ours     &  \textbf{0.64} & \textbf{1.66} & \textbf{3.76}  \\\hline
\end{tabular}
\end{table}

\subsection{Limitation and discussion}\label{sec:limitation}

Although our model can achieve stable and efficient photo enhancement across a variety of scenes, it has two limitations inherited from 3D LUT. First, our 3D LUT can adapt to different images based on their content; however, once the 3D LUT is determined for an input image, it is the same for different local areas within the image. Consequently, 3D LUT may produce less satisfactory results in some areas requiring local enhancement. Fig. \ref{figure:limitation_localcontrast} shows such an example, where the scene of input image has very high dynamic range. As can be seen, although our model properly enhances the global color and tone of the input image, the local contrast in the shadow area is not sufficiently enhanced.

One possible solution to address this limitation is to combine some local contrast enhancement method, such as the local tone mapping algorithm \cite{durand2002fast}, with our 3D LUT. Here we give one example by replacing the bilateral filter used in \cite{durand2002fast} with the guided filter in \cite{he2012guided} for its better performance on edge preserving. We first apply local tone mapping to the input image and then apply the proposed 3D LUT to the tone mapped output. As can be seen from Fig. \ref{figure:limitation_localcontrast}, combining local tone mapping with our 3D LUT achieves good visual quality on both global and local enhancement in this challenging case. However, it should be noted that the local tone mapping method is still time consuming for high resolution images, and how to efficiently embed local enhancement operations into 3D LUT will be our future work. 

Second, as a compact and highly efficient operator, 3D LUT transforms each input RGB value independently, which however limits its capability on detail enhancement. One challenging example is shown in Fig. \ref{figure:limitation_noise}, where the input image was taken at nighttime with heavy noise. Compared to the result generated by Google's image processing pipeline \cite{hasinoff2016burst}, where there are some denoising operations, our 3D LUT model effectively enhances the dark areas but also amplifies the noise in these areas. In practice, one possible solution to address this limitation is to combine 3D LUT transformation with some denoising module as commonly done in camera imaging pipeline \cite{karaimer2016software,hasinoff2016burst}. We also leave this for our future work. 

\section{Conclusion}

We proposed an effective and efficient learning method for image-adaptive photo enhancement. Several basis 3D LUTs were learned in couple with a lightweight CNN weight predictor so that for each input image, an adaptive 3D LUT could be generated by fusing the basis 3D LUTs according to the image content. Smooth and monotonicity regularization terms were introduced into the learning process to make our model stable and robust to various scenes. We evaluated the proposed method on two photo enhancement applications, photo retouching and imaging pipeline enhancement, under both the pairwise learning and unpaired learning settings. The results demonstrated that our model consistently outperforms state-of-the-art deep image enhancement methods in terms of both quantitative and qualitative comparisons. Particularly, our model is more than 100 times faster than previous ones on processing images with 4K or higher resolutions, and it consumes much lower memory. It provides a promising solution to practical, especially high-resolution, photo enhancement applications.



\bibliographystyle{IEEETran}
\bibliography{egbib}

\begin{IEEEbiography}[{\includegraphics[width=0.9in,height=1.0in,keepaspectratio]{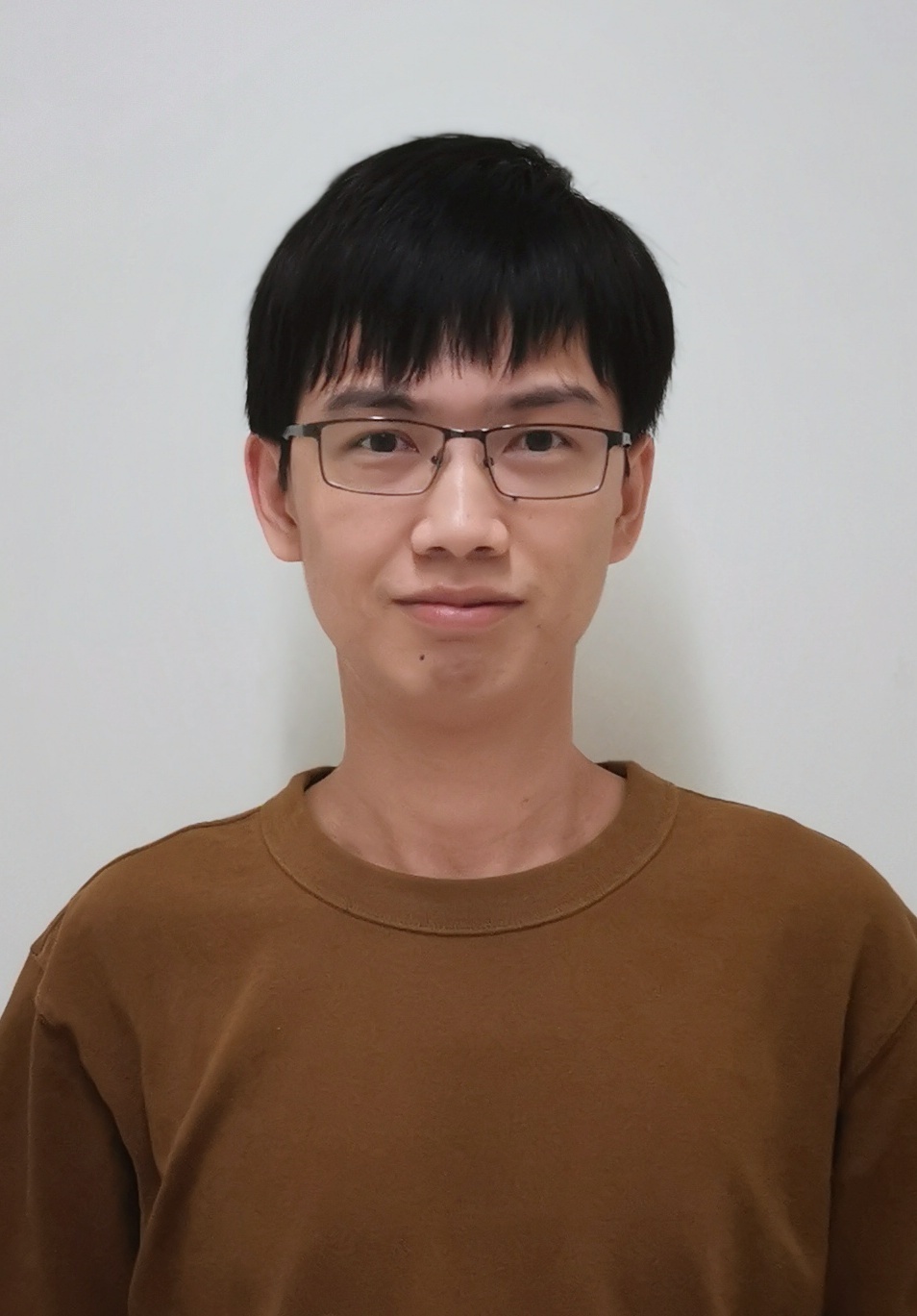}}]{$\\$Hui Zeng} is currently a Ph.D candidate in the Department of Computing, The Hong Kong Polytechnic University. He received his B.Sc. and M.Sc. degrees from the Faculty of Electronic Information and Electrical Engineering, Dalian University of Technology, in 2014 and 2016, respectively. His research interests include image processing and computer vision.
\end{IEEEbiography}
\begin{IEEEbiography}[{\includegraphics[width=0.9in,height=1.0in,keepaspectratio]{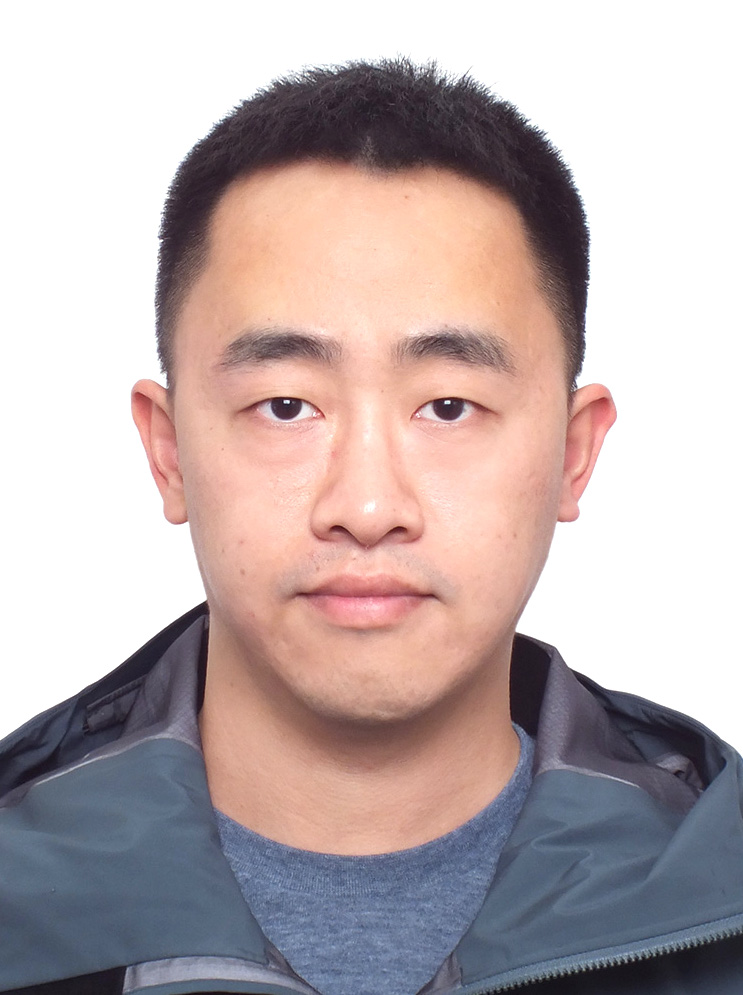}}]{$\\$Jianrui Cai} is currently a Ph.D candidate in the Department of Computing, The Hong Kong Polytechnic University. He received his B.Sc. and M.Sc. degrees from the College of Computer Science and Electronic Engineering, Hunan University, China in 2012 and 2015, respectively. His research interests include image processing, computational photography and computer vision.
\end{IEEEbiography}
\begin{IEEEbiography}[{\includegraphics[width=0.9in,height=1.0in,keepaspectratio]{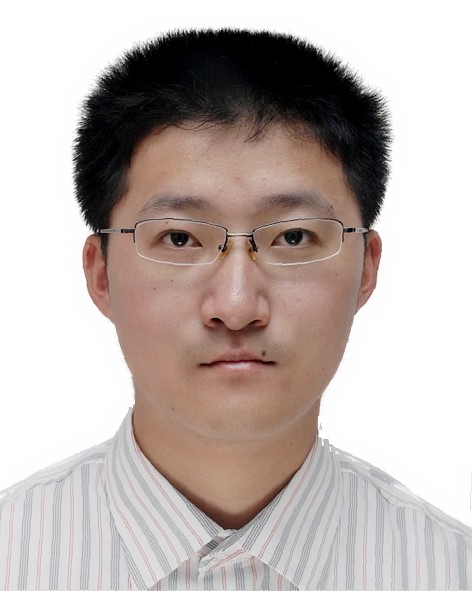}}]{$\\$Lida Li} received the B.S. and M.Sc. degrees from the School of Software Engineering, Tongji University,
Shanghai, China, in 2013 and 2016, respectively. He is now pursuing his Ph.D. degree at the Department of Computing, The Hong Kong Polytechnic University. His research interests are machine learning and face analysis.
\end{IEEEbiography}
\begin{IEEEbiography}[{\includegraphics[width=0.9in,height=1.0in,keepaspectratio]{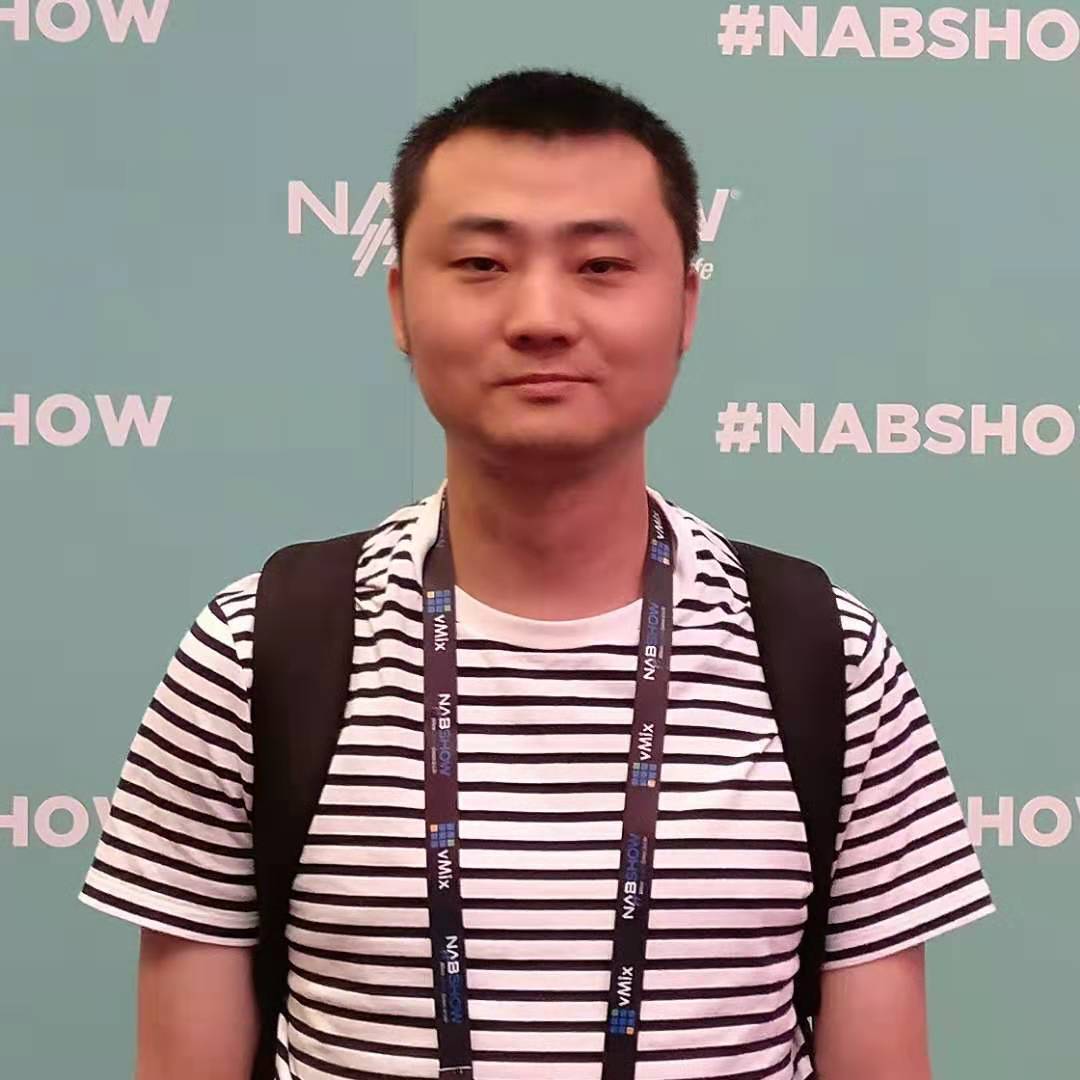}}]{$\\$Zisheng Cao} is currently with imaging group of DJI. He received his B.S. and M.S. degrees from Tsinghua University, in 2005 and 2007, respectively, and PhD degree from the University of Hong Kong, in 2014. His research interests include image signal processing and machine learning. Before joining in DJI, he was a research scientist in Philips.
\end{IEEEbiography}
\begin{IEEEbiography}[{\includegraphics[width=0.9in,height=1.0in,clip,keepaspectratio]{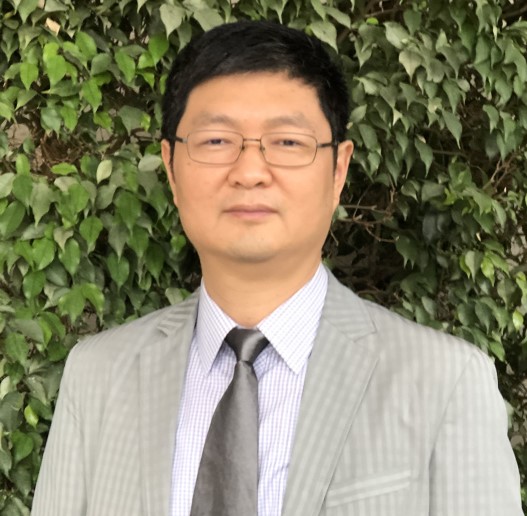}}]{$\\$Lei Zhang}
(M’04, SM’14, F’18) received his B.Sc. degree in 1995 from Shenyang Institute of Aeronautical Engineering, Shenyang, P.R. China, and M.Sc. and Ph.D degrees in Control Theory and Engineering from Northwestern Polytechnical University, Xi’an, P.R. China, in 1998 and 2001, respectively. From 2001 to 2002, he was a research associate in the Department of Computing, The Hong Kong Polytechnic University. From January 2003 to January 2006 he worked as a Postdoctoral Fellow in the Department of Electrical and Computer Engineering, McMaster University, Canada. In 2006, he joined the Department of Computing, The Hong Kong Polytechnic University, as an Assistant Professor. Since July 2017, he has been a Chair Professor in the same department. His research interests include Computer Vision, Image and Video Analysis, Pattern Recognition, and Biometrics, etc. Prof. Zhang has published more than 200 papers in those areas. As of 2020, his publications have been cited more than 54,000 times in literature. Prof. Zhang is a Senior Associate Editor of IEEE Trans. on Image Processing, and is/was an Associate Editor of IEEE Trans. on Pattern Analysis and Machine Intelligence, SIAM Journal of Imaging Sciences, IEEE Trans. on CSVT, and Image and Vision Computing, etc. He is a ``Clarivate Analytics Highly Cited Researcher" from 2015 to 2019. More information can be found in his homepage \url{http://www4.comp.polyu.edu.hk/~cslzhang/}.
\end{IEEEbiography}




\end{document}